%% file: main.tex
\documentclass[letterpaper,twocolumn,10pt]{article}
\usepackage{usenix-2020-09}
\usepackage{tikz}
\usepackage{amsmath}
\usepackage{amsmath,amssymb,amsfonts}
\usepackage{soul}
\usepackage{algorithmic}
\usepackage{graphicx}
\usepackage{textcomp}
\usepackage{xcolor}
\usepackage{graphicx}
\usepackage{balance}  
\usepackage{booktabs} 

\usepackage{enumitem}
\usepackage{graphicx}
\usepackage{algorithmic}
\usepackage{array}
\usepackage{mdwmath}
\usepackage{mdwtab}
\usepackage{eqparbox}
\usepackage[tight,footnotesize]{subfigure}
\usepackage{caption}
\usepackage{stfloats}
\usepackage{url}

\newcolumntype{P}[1]{>{\centering\arraybackslash}p{#1}}

\usepackage{url}
\usepackage{algorithmic}
\usepackage{algorithm2e}
\usepackage{array}
\usepackage{multirow}
\newcolumntype{P}[1]{>{\centering\arraybackslash}p{#1}}

\usepackage{verbatim}
\usepackage{amsmath}
\usepackage{amsfonts}
\usepackage{amssymb}
\usepackage{tabularx}
\usepackage{caption}

\usepackage{epstopdf}
\usepackage{notoccite}
\usepackage{stackengine}

\usepackage{afterpage}

\usepackage{slashbox}
\usepackage{authblk}

\begin{document}

\date{}

\title{\Large \bf Are Your Sensitive Attributes Private?\\
Novel Model Inversion Attribute Inference Attacks on Classification Models}

\author[1]{Shagufta Mehnaz}
\author[1]{Sayanton V. Dibbo}
\author[1]{Ehsanul Kabir}
\author[2]{Ninghui Li}
\author[2]{Elisa Bertino}
\affil[1]{Department of Computer Science, Dartmouth College}
\affil[2]{Department of Computer Science, Purdue University}
\affil[ ]{\textit {\{shagufta.mehnaz, sayanton.v.dibbo.gr, ehsanul.kabir.gr\}@dartmouth.edu, \{ninghui, bertino\}@purdue.edu}}

\maketitle

\begin{abstract}
Increasing use of machine learning (ML) technologies in privacy-sensitive domains such as medical diagnoses, lifestyle predictions, and business decisions highlights the need to better understand if these ML technologies are introducing leakage of sensitive and proprietary training data. In this paper, we focus on model inversion attacks where the adversary knows non-sensitive attributes about records in the training data and aims to infer the value of a sensitive attribute unknown to the adversary, using only black-box access to the target classification model. We first devise a novel confidence score-based model inversion attribute inference attack that significantly outperforms the state-of-the-art. We then introduce a label-only model inversion attack that relies only on the model's predicted labels  but still matches our confidence score-based attack in terms of attack effectiveness. We also extend our attacks to the scenario where some of the other (non-sensitive) attributes of a target record are unknown to the adversary. We evaluate our attacks on two types of machine learning models, decision tree and deep neural network, trained on three real datasets. Moreover, we empirically demonstrate the disparate vulnerability of model inversion attacks, i.e., specific groups in the training dataset (grouped by gender, race, etc.) could be more vulnerable to model inversion attacks. 
\end{abstract}

\input{Sections/I_Introduction}
\input{Sections/II_ProblemDefinition}

\input{Sections/III_Metrics}
\input{Sections/IV_NewAttack}

\input{Sections/V_NewAttacksPartial}

\input{Sections/VI_Evaluations}

\input{Sections/VII_RelatedWork}

\input{Sections/VIII_Conclusions}

\input{Sections/Bib-Appendix}

\end{document}

%% file: Sections/I_Introduction.tex
\vspace{-0.2cm}
\section{Introduction}
\label{sec:intro}
\vspace{-0.3cm}
Across numerous sectors, the use of ML technologies trained on proprietary and sensitive datasets has increased significantly, 
e.g., in the domains of personalized medicine \cite{medicine1, medicine2, medicine3, medicine4}, product recommendation~\cite{recom1, recom2, recom3}, finance and law~\cite{finlaw1, finlaw2, finlaw3}, social media~\cite{social1, social2, social3}, etc.
{Companies provide access to such trained ML models through APIs whereas users querying these models are charged on a pay-per-query basis.}
With the increasing use of ML technologies in personal data, we have seen a recent surge of serious privacy concerns that were previously ignored~\cite{papernot2018, shokri2017, tramer2016stealing, FredriksonCCS2015}. Therefore, it is  important to investigate whether public access to such trained models introduces new attack vectors against the privacy of these proprietary and sensitive datasets used for training ML models.
A \emph{model inversion attack} is one of such attacks on ML models  that turns the one-way journey from training data to model into a two-way one, i.e., this attack allows an adversary to infer part of the training data when it is given access to the target ML model.

Fredrikson et al.~\cite{FredriksonCCS2015, PharmaUSENIX2014} proposed \emph{two} formulations of model inversion attacks. In the first one, which we call \textbf{model inversion attribute inference (MIAI)} attack, the adversary aims to learn some sensitive attribute of an individual whose data are used to train the target model, and whose other attributes are known to the adversary.  This can be applied, e.g., when each instance gives information about one individual.  In the second formulation, which we call \textbf{typical instance reconstruction (TIR)} attack, the adversary is given access to a classification model and a particular class, and aims to come up with a typical instance for that class.  For example, the adversary, when given access to a model that recognizes different individuals' faces, tries to reconstruct an image that is similar to a target individual's actual facial image. 

Several recent studies investigate TIR attacks~\cite{avodji2019gamin,YangCCS2019,Zhang_2020_CVPR}.  For TIR attacks to be considered successful, it is not necessary for a reconstructed instance to be quantitatively close to any specific training instance. In contrast, MIAI attacks are evaluated by the ability to predict \emph{exact} attribute values of individual instances. Evaluation of TIR attacks is typically done by having humans assessing the similarity of the reconstructed instances (e.g., reconstructed facial images) to training instances.  Thus a model that is able to learn the essence of each class and generalizes well (as opposed to relying on remembering information specific to training instances) will likely remain vulnerable to such an attack.  Indeed, it has been proven~\cite{Zhang_2020_CVPR} that a model's predictive power and its vulnerability to such TIR attacks are two sides of the same coin. This is because highly predictive models are able to establish a strong correlation between features and labels and this is the property that an adversary exploits to mount the TIR attacks~\cite{Zhang_2020_CVPR}.  In other words, the existence of TIR attacks is a feature of good classification models, although the feature may be  undesirable in some settings. We investigate whether the root cause of TIR attacks (high predictive power) also applies to MIAI attacks. According to our observation, we point out that such is not the case.

In this paper, we focus only on MIAI attacks on classification models where data about individuals are used.  More specifically, we consider the attribute inference attacks where the adversary leverages \emph{black-box} access to an ML model to infer the sensitive attributes of a target individual.  While attribute inference in other contexts has been studied extensively in the privacy literature (e.g., user attribute inference in  social networks~\cite{GongTOPS2018, AttriInferWWW2017}), there exists little work studying to what extent model inversion introduces new attribute inference vulnerabilities. In the rest of the paper, we refer to MIAI attacks whenever we use the term model inversion attack.

\textbf{Proposed new model inversion attacks:} 
In this paper, we devise two new black-box model inversion attribute inference (MIAI) attacks: (1) confidence score-based MIAI attack (CSMIA) and (2) label-only MIAI attack (LOMIA). 
The confidence score-based MIAI attack assumes that the adversary has access to the target model's confidence scores whereas the label-only MIAI attack assumes the adversary's access to the target model's label predictions only. 
To the best of our knowledge, ours is the first work to propose a label-only MIAI attack.
We empirically show that despite having access to only the predicted labels, our label-only attack performs on par with the proposed confidence score-based attack.
Also, both of our proposed attacks outperform state-of-the-art attacks significantly. Furthermore, we note that defense mechanisms~\cite{FredriksonCCS2015} that reduce the precision of confidence scores or introduce noise in the confidence scores to thwart model inversion attacks are ineffective against our label-only attack.

While the existing attacks~\cite{FredriksonCCS2015, PharmaUSENIX2014} assume that the adversary has full knowledge of other non-sensitive attributes of the target record, it is not clear how the adversary would perform in a setting where it has only partial knowledge of those attributes. To understand the vulnerability of model inversion attacks in such practical scenarios, we also propose {extensions of our attacks that work}
even when some non-sensitive attributes are unknown to the adversary. Moreover, we also investigate if there are scenarios when  model inversion attacks do not threaten the privacy of the overall dataset but are effective on some specific groups of instances (e.g., records grouped by race, gender, occupation, etc.).
We empirically show that there exists such discrimination across different groups of the training dataset where a group is more vulnerable than the others. We use the term \emph{disparate vulnerability} to represent such discrimination.
We further investigate if model inversion attribute inference attacks are  able to infer the sensitive attributes in data records that do not belong to the training dataset of the target model but are drawn from the same distribution. A model inversion attack with such capability compromises the privacy of not only the target model's training dataset but also breaches its \emph{distributional privacy}.

We train two models-- a \emph{decision tree} and a \emph{deep neural network} with each of the {three} real datasets in our experiments, General Social Survey (GSS)~\cite{gss}, Adult dataset~\cite{adult}, {and FiveThirtyEight dataset~\cite{538}}, to evaluate our proposed attacks. 
To the best of our knowledge, ours is the first work that studies MIAI attacks  in such details on \emph{tabular} datasets which is the most common data type used in real-world ML~\cite{luo2020network}. 

\textbf{Effective evaluation of model inversion attacks:} 
Although the  Fredrikson  et al.  attack~\cite{FredriksonCCS2015}  primarily  uses  accuracy to evaluate model inversion attacks, in this paper, we argue that accuracy is not the best measure. This is because simply predicting the majority class for all the instances can achieve very high accuracy which certainly misrepresents the performances of model inversion attacks. Moreover, we argue that the F1 score, a widely used metric, is also not sufficient by itself since it emphasizes only the positive class, and simply predicting the positive class for all the instances can achieve a significant F1 score. Hence, we propose to also use G-mean~\cite{Sun2009ClassificationOI} and Matthews correlation coefficient (MCC)~\cite{MATTHEWS1975442} as metrics in addition to precision, recall, accuracy, {false positive rate (FPR)}, and F1 score to design a framework that can effectively evaluate any model inversion attack.

{
While the existing MIAI attacks~\cite{FredriksonCCS2015, PharmaUSENIX2014} evaluate their performance on binary sensitive attributes only, we evaluate our attacks on multi-valued sensitive attributes as well. We use attack confusion matrices to evaluate the attack performances in estimating multi-valued sensitive attributes. Moreover, we evaluate cases where an adversary aims to estimate multiple sensitive attributes of a target record which also has not been explored in the existing MIAI attacks~\cite{FredriksonCCS2015, PharmaUSENIX2014}. Finally, we evaluate the number of queries to the black-box target models to perform the proposed attacks.
}

\textbf{Comparison with baseline attribute inference attacks}: We also compare the performances of various model inversion attacks with those from attacks that do not query the target model, e.g., randomly guessing the sensitive attribute according to some distribution.
When a particular model inversion attack deployed against a target model performs similarly to such attacks, we can conclude that the target model in not vulnerable to that particular model inversion attack. Hence, in this paper, we address the following general research question- is it possible to identify when a model should be classified as vulnerable to such model inversion attacks? More specifically, does black-box access to a particular model really help the adversary to estimate the sensitive attributes 
which is otherwise impossible for the adversary?
We demonstrate that our proposed attacks significantly outperform baseline attribute inference attacks that do not require access to the target model.

\begin{table*}[h]
  \centering
  \caption{Assumption of adversary capabilities/knowledge for different attack strategies.}
  \vspace{-0.2cm}
  \resizebox{1\textwidth}{!}{
  \begin{tabular}{ | l | c | c | c | c | c | c | c |}
    \hline
    \multirow{3}{*}{Attack strategy} & Predicted   & Confidence score & Target individuals' all & All possible & Marginal prior of & Marginal prior  & Confusion     \\ 
        & label  & along with & non-sensitive attributes  & values of the & the sensitive    & of all other (non-  & matrix of \\ 
        &  & predicted label & {including true label} & sensitive attribute & attribute & sensitive) attributes & the model  \\ \hline 
    NaiveA     &     &   &      &\checkmark &\checkmark &   &      \\ \hline
    RandGA   &    &    &     &\checkmark &\checkmark(optional) &   &        \\ \hline
    FJRMIA~\cite{FredriksonCCS2015} & \checkmark &  & \checkmark &\checkmark &  \checkmark & \checkmark  &  \checkmark \\ \hline
    CSMIA & \checkmark & \checkmark   &\checkmark &\checkmark &  &   & \\ \hline
    LOMIA & \checkmark &    &\checkmark &\checkmark &  &   & {\checkmark} \\ \hline
  \end{tabular}
  }
  \vspace{-0.5cm}
\label{table:adversary_assumptions}
\end{table*}

\vspace{0.1cm}
\textbf{Summary of contributions}: In summary, this paper makes the following contributions:

\begin{enumerate}
\vspace{-0.2cm}
    \item We design two new black-box model inversion attribute inference (MIAI) attacks: (1) confidence score-based MIAI attack and (2) label-only MIAI attack. We define the various capabilities of the adversary and provide a detailed threat model.
    \item  We conduct an extensive evaluation of our attacks using two types of ML models, decision tree and deep neural network, trained with {three} real datasets.  Evaluation results show that our proposed attacks significantly outperform the existing attacks. Moreover, our label-only attack performs on par with the proposed confidence score-based MIAI attack despite having access to only the predicted labels of the target model.
    \item We extend { both of our proposed attacks} to the scenario where some of the other (non-sensitive) attributes of a target record are unknown to the adversary and demonstrate that the performance of our attacks is not impacted significantly in those circumstances.
    \item We  uncover that a particular subset of the training dataset (grouped by attributes, such as gender, race, etc.) could be more vulnerable than others to the model inversion attacks, a phenomenon we call \emph{disparate vulnerability}.
\end{enumerate}

%% file: Sections/II_ProblemDefinition.tex
\section{Problem Definition and Existing Attacks}
\vspace{-0.1cm}
\subsection{Model Inversion Attribute Inference}
\label{subsec:mia}
\vspace{-0.1cm}
An ML model can be represented using a deterministic function $f$
where the input of this function is a $d$-dimensional vector $\mathbf{x} = [x_1, x_2, ..., x_d]$
that represents $d$ attributes and $y' \in \mathcal{Y}$ is the output. In the case of a regression problem, $\mathcal{Y} = \mathcal{R}$. However, in this work, we focus on classification problems. Therefore, more specifically, $f$
outputs $y'$ if it returns only the predicted label and outputs $\mathcal{R}^m$ if it returns the confidence scores as well, where $m$ is the number of unique class labels ($y_1, y_2, ..., y_m$) and $\mathcal{R}^m$ represents the confidence scores returned for these $m$ class labels. Finally, the class label with the highest confidence score is considered as the output of the prediction model. We denote the dataset on which the $f$ model is trained as $DS_T$.
From now on, we use the term $y$ to represent the actual value in the training dataset $DS_T$ whereas $y'$ is used to represent the model output $f(\mathbf{x})$. The values of $y$ and $y'$ are the same in the case of a correct prediction or different in the case of an incorrect prediction by  $f$.

Now, some of the attributes in $\mathbf{x}$ introduced above could be privacy sensitive. Without loss of generality, let's assume that $x_1 \in \mathbf{x}$ is a sensitive attribute
that the individual corresponding to a data record in the training dataset does not want to reveal to the public. However, a model inversion attack may allow an adversary to infer this $x_1$ attribute value of a target individual given some specific capabilities, such as, access to the black-box model (i.e., target model), background knowledge about the target individual, etc.

\subsection{Threat Model}
\label{subsec:threat_model}
\vspace{-0.1cm}
The adversary is assumed to have all or a subset of the following capabilities/knowledge (see Table~\ref{table:adversary_assumptions}):
\begin{itemize}
    \vspace{-0.15cm}
    \item Access to the black-box target model, i.e., the adversary can query the model with $\mathbf{x} = [x_1, x_2, ..., x_d]$ and obtain a class label $y'$ as the output. 
    \vspace{-0.15cm}
    \item The confidence scores returned by the target model for $m$ class labels, i.e.,  $\mathcal{R}^m$. 
    \vspace{-0.15cm}
    \item {Full/partial knowledge of the non-sensitive attributes and also knowledge of the true label of the target record.}
    \vspace{-0.15cm}
    \item All possible ($k$) values of the sensitive attribute $x_1$.
    \vspace{-0.15cm}
    \item Knowledge of marginal prior of the sensitive attribute $x_1$, i.e., $\mathbf{p_1} = \{p_{1,1}, p_{1,2}, ..., p_{1,k}\}$ where $k$ is the number of all possible values of $x_1$ and $p_{1,k}$ is the probability of the $k-$th unique possible value. 
    \vspace{-0.15cm}
    \item Knowledge of confusion matrix ($\mathcal{C}$) of the model where $\mathcal{C}[y, y']$ = $Pr[f(x) = y' | y$ is the true label$]$. {Here, confusion matrix represents the performance of an ML model when queried on the entire training dataset~\cite{FredriksonCCS2015}.}
    \vspace{-0.15cm}
\end{itemize}

Note that, for the attacks designed in this paper, the adversary does not need the knowledge of marginal priors of any attributes (sensitive or non-sensitive). {While our CSMIA strategy does not require the knowledge of target model confusion matrix, the LOMIA strategy indirectly assumes this knowledge.}
The adversary has only black-box access to the model, i.e., it has no knowledge of the model details (e.g., architecture or parameters). Finally, we only consider a passive adversary that does not aim to corrupt the machine learning model or influence its output in any way.

\vspace{-0.15cm}
\subsection{Baseline Attack Strategies}
\label{subsec:existing_attacks}
\vspace{-0.1cm}
\subsubsection{Naive Attack (NaiveA)}
\vspace{-0.15cm}
A naive model inversion attack assumes that the adversary has knowledge about the probability distribution (i.e., marginal prior) of the sensitive attribute and always predicts the sensitive attribute to be the value with the highest marginal prior. Therefore, this attack does not require access to the target model. 
Note that this attack can still achieve significant accuracy if the sensitive attribute is highly unbalanced, e.g., if the sensitive attribute can take only two values and there is an 80\%-20\% probability distribution, predicting the value with higher probability would result in 80\% accuracy.

\vspace{-0.1cm}
\subsubsection{Random Guessing Attack (RandGA)}
\vspace{-0.15cm}
The adversary in this attack also does not require access to the target model. The adversary randomly predicts the sensitive attribute by setting a probability for each possible value. The adversary may or may not have access to the marginal priors of the sensitive attribute. Fig.~\ref{fig:random_30} in Appendix~\ref{appn:random} shows the optimal performance of random guessing attack in terms of different metrics when the adversary sets different probabilities for predicting the positive class sensitive attribute is independent of its knowledge of marginal prior ($0.3$ in this example). Note that, predicting the positive class for all the instances with this attack (i.e., setting a probability 1 for the positive class) would result in a significantly high F1 score, mainly due to a recall of $100\%$ (Fig.~\ref{fig:random_30} in Appendix).

\subsection{Fredrikson et al. Attack~\cite{FredriksonCCS2015} (FJRMIA)}
\label{subsec:ccs15}
\vspace{-0.15cm}
The Fredrikson et al.~\cite{FredriksonCCS2015} black-box model inversion attack assumes that the adversary can obtain the model's predicted label, has knowledge of all the attributes of a targeted record (including the true $y$ value) except the sensitive attribute, has access to the marginal priors of all the attributes, and also to the confusion matrix of the target model (see Table~\ref{table:adversary_assumptions}). The adversary can query the target model multiple times by varying the sensitive attribute ($x_1$) and obtain the predicted $y'$ values.
After querying the model $k$  times with $k$ different $x_1$ values ($x_{1,0}, x_{1,1}, \ldots, x_{1,k-1}$) while keeping the other known attributes unchanged, the adversary computes $\mathcal{C}[y, y'] * p_{1,i}$ for each possible sensitive attribute value, where
\begin{equation}
\footnotesize
{
\mathcal{C}[y, y'] = Pr[f(x) = y' |\;y\;is\;the\;true\;label] \nonumber
}
\end{equation}
and $p_{1,i}$ is the marginal prior of i-th possible sensitive attribute value. Finally, the attack predicts the sensitive attribute value for which the  $\mathcal{C}[y, y'] * p_{1,i}$ value is the maximum.

%% file: Sections/III_Metrics.tex
\vspace{-0.15cm}
\section{Metrics for Evaluating MIAI Vulnerability }
\label{sec:indepth}
\vspace{-0.15cm}
Though the impact of model inversion attacks can be overwhelming, in this section, we aim to take a deep dive to understand if it is possible to determine when a model should be classified as vulnerable and if the metrics considered in the existing model inversion attack research are sufficient. More specifically, we investigate the following general research question- \emph{does black-box access to a particular model really help the adversary to estimate the sensitive attributes which is otherwise impossible for the adversary to estimate (i.e., without access to that black-box model)?} 

Understanding a model's vulnerability to inversion attacks requires a meaningful metric to evaluate and compare different model inversion attacks. The FJRMIA~\cite{FredriksonCCS2015} primarily uses accuracy. However, if we care only about accuracy, the naive attack of simply guessing the majority class for all the instances can achieve very high accuracy. Another widely used metric is the F1 score. However, the F1 score of the positive class emphasizes only on that specific class and thus, as a one-sided evaluation, cannot be considered as the \emph{only metric} to evaluate the attacks. Otherwise, always guessing the positive class may achieve similar or even better F1 score (mainly due to a recall of $100\%$) than any sophisticated model inversions attack that identifies the positive class instances more strategically.
Therefore, to understand whether access to the black-box model considerably contributes to attack performance and also to compare the baseline attack strategies (that do not require access to the model, i.e., naive attack and random guessing attack) to our proposed attacks, we use the following two metrics in addition to precision, recall, accuracy, {FPR,} and F1 score: G-mean~\cite{Sun2009ClassificationOI} and Matthews correlation coefficient (MCC)~\cite{MATTHEWS1975442}, as described below.

\textbf{G-mean}: G-mean is the geometric mean of sensitivity and specificity~\cite{Sun2009ClassificationOI}. Thus it takes all of true positives (TP), true negatives (TN), false positives (FP), and false negatives (FN) into account. With this metric, the random guessing attack can achieve a maximum performance of $50\%$. Note that, even if the adversary has knowledge of marginal priors of the sensitive attribute, it is not able to achieve a G-mean value of more than $50\%$ by setting different probabilities for predicting the positive class sensitive attribute (Fig.~\ref{fig:random_30} in Appendix). For  random guessing attack, the optimal G-mean value can be achieved by setting the probability to $0.5$. The G-mean for the naive attack is always $0\%$.
\vspace{-0.1cm}
\begin{equation}
\footnotesize
{
G-mean = \sqrt{\frac{TP}{TP + FN} * \frac{TN}{TN + FP}}
}
\end{equation}
\textbf{Matthews correlation coefficient (MCC)}: This MCC metric also takes into account all of TP, TN, FP, and FN, and is a balanced measure which can be used even if the classes of the sensitive attribute are of very different sizes~\cite{MATTHEWS1975442}. It returns a value between -1 and +1. A coefficient of +1 represents a perfect prediction, 0 represents a prediction no better than the random one, and -1 represents a prediction that is always incorrect. Note that, even if the adversary has the knowledge of marginal priors of the sensitive attribute, it is not able to achieve an MCC value of more than $0$ with the random guessing attack strategy (details in Appendix~\ref{appn:random}). Also, the naive attack always results in an MCC of $0$, independent of the marginal prior knowledge (either TP=FP=0 or TN=FN=0). 
\begin{equation}
\footnotesize
{
MCC = \frac{(TP*TN)-(FP*FN)}{\sqrt{(TP+FP)*(TP+FN)*(TN+FP)*(TN+FN)}}
}
\end{equation}

%% file: Sections/IV_NewAttack.tex
\section{New Model Inversion Attacks}
\label{sec:new_attacks}
\vspace{-0.2cm}
We design two new attack strategies: 
(1) confidence score-based model inversion attack (CSMIA) and 
(2) label-only model inversion attack (LOMIA).
Table~\ref{table:adversary_assumptions} shows the different adversary capabilities/knowledge assumptions for these attacks in contrast to the existing attacks.

\begin{figure*}[h]
\centering
\includegraphics[width=0.99\textwidth, height=3.2cm]
{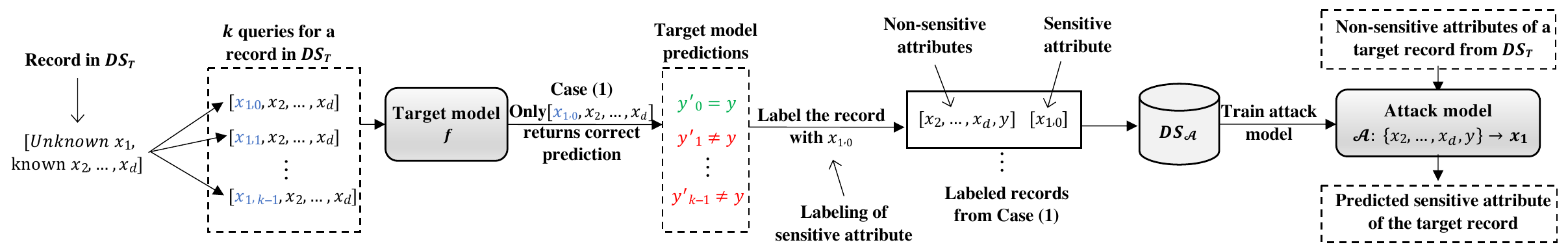}
\caption{Label-only model inversion attack (LOMIA). First, the adversary collects the case (1) records by querying the target model $f$, obtains the $DS_{\mathcal{A}}$ dataset, and trains the attack model $\mathcal{A}$. The adversary then leverages the trained attack model to predict the sensitive attribute values of the target records.}
\vspace{-0.2cm}
\label{fig:lomia}
\end{figure*}

\subsection{Confidence Score-based Model Inversion Attack (CSMIA)} 
\label{subsec:cscore_attack}
\vspace{-0.2cm}
This attack exploits the confidence scores returned by the target model. Unlike FJRMIA~\cite{FredriksonCCS2015}, the adversary assumed in this attack does not have access to the marginal priors or the confusion matrix. 
{The adversary knows the true labels for the records it is attacking (Table~\ref{table:adversary_assumptions}).}
The \emph{key idea} of this attack is that the target model's returned prediction is more likely to be correct and the confidence score is more likely to be higher when it is queried with a record containing the original sensitive attribute value (since the target model encountered the target record with original sensitive attribute value during training). In contrast, the target model's returned prediction is more likely to be incorrect when it is queried with a record containing the wrong sensitive attribute value.

The adversary first queries the model by setting the sensitive attribute value $x_1$ to all possible $k$ values while all other known input attributes of the target record remain the same. {If the sensitive attribute is continuous, we can use binning to turn it into a categorical attribute and recover an approximate value.}
If there are two possible values of a sensitive attribute (i.e., $k$ = 2, well depicted by a yes/no answer from an individual in response to a survey question), the adversary queries the model by setting the sensitive attribute value $x_1$ to both \emph{yes} and \emph{no} while all other known input attributes of the target record remain the same. Let $y'_{0}$ and $conf_0$ be the returned model prediction and confidence score when the sensitive attribute is set to  \emph{no}. Similarly, $y'_{1}$ and $conf_1$ are the model prediction and confidence score when the sensitive attribute is set to \emph{yes}. 
In order to determine the value of $x_1$, this attack considers the following three cases:
    \begin{enumerate}[label=\fbox{\textbf{Case (\arabic*)}},  leftmargin=0pt, itemindent=53pt]
        \item If the target model's prediction is correct \emph{only for a single sensitive attribute value}, e.g., $y$ = $y'_{0}$ $\wedge$ $y$ != $y'_{1}$ or $y$ != $y'_{0}$ $\wedge$ $y$ = $y'_{1}$ in the event of a binary sensitive attribute, the attack selects the sensitive attribute to be the one for which the prediction is correct. For instance, if $y$ = $y'_{1}$ $\wedge$ $y$ != $y'_{0}$, the attack predicts $yes$ for the sensitive attribute and vice versa. \emph{Note that, for this case, the adversary only requires the predicted labels and does not require  the confidence scores}. We leverage the records that fall into this case in our label-only model inversion attack as described later in Section~\ref{subsec:labelonly_attack}.
        \vspace{-0.2cm}
        \item If the model's prediction is correct for multiple sensitive attribute values, i.e., $y$ = $y'_{0}$ $\wedge$ $y$ = $y'_{1}$, the attack selects the sensitive attribute to be the one for which the prediction confidence score is the maximum. In the above example, if the model's prediction is correct with higher confidence when $yes$ value is set for the sensitive attribute, the attack outputs the $yes$ value for the $x_1$ prediction and vice versa. 
        \vspace{-0.2cm}
        \item If the model outputs incorrect predictions for all possible sensitive attribute values, i.e., $y$ != $y'_{0}$ $\wedge$ $y$ != $y'_{1}$, the attack selects the sensitive attribute to be the one for which the prediction confidence is the minimum. In the above example, if the model outputs the incorrect prediction with higher confidence when $yes$ value is set for the sensitive attribute, the attack predicts the $no$ value for $x_1$ and vice versa. 
    \end{enumerate}

\vspace{-0.1cm}
\subsection{Label-Only Model Inversion Attack (LOMIA)} 
\label{subsec:labelonly_attack}
\vspace{-0.2cm}
This advanced attack assumes the adversary's access to the target model's predicted labels only. Therefore, defense mechanisms~\cite{FredriksonCCS2015} that reduce the precision of confidence scores or introduce noise in the confidence scores in order to thwart model inversion attacks are ineffective against our label-only attack. The attack has the following steps as shown in Figure~\ref{fig:lomia}: (1) obtaining an attack dataset ($DS_{\mathcal{A}}$), (2) training an attack model $\mathcal{A}$ from $DS_{\mathcal{A}}$, and (3) leveraging $\mathcal{A}$ to infer the sensitive attributes of target records. 

\vspace{-0.1cm}
\subsubsection{Obtaining Attack Dataset $DS_{\mathcal{A}}$} 
\label{subsec:lomia_dsa}
\vspace{-0.1cm}
The \emph{key intuition} of this attack step is that if the target model $f$ returns the correct prediction ($y$) for only one possible value of the sensitive attribute, it is highly likely that this particular value represents the original sensitive attribute value, e.g., sensitive attribute value $x_{1,0}$ in Figure~\ref{fig:lomia}. Hence, the adversary then labels the record in this example with $x_{1,0}$. The adversary collects all such labeled records that fall into Case (1) as described in Section~\ref{subsec:cscore_attack} and obtains the $DS_{\mathcal{A}}$ dataset. Note that, the labeling of sensitive attributes might have some errors, e.g., $x_{1,0}$ in Figure~\ref{fig:lomia} might not be the original sensitive attribute of the record even though only with this value the target model returned the correct prediction. 
Table~\ref{table:dsa} in Section~\ref{subsec:expres_model} shows the sizes of the $DS_{\mathcal{A}}$ datasets obtained from different target models in our experiments and their corresponding accuracy. However, since the LOMIA attacker does not know the original sensitive attribute values, it uses the entire $DS_{\mathcal{A}}$ datasets to train the attack models.

{Note that, while building the attack model dataset $DS_{\mathcal{A}}$, we assume that the adversary knows the real $y$ attribute of all the instances in the training dataset. In other words, unlike CSMIA, the adversary in LOMIA strategy assumes the knowledge of the target model confusion matrix (Table~\ref{table:adversary_assumptions}). 
}

\subsubsection{Training Attack Model $\mathcal{A}$} 
\label{subsec:lomia_A}
\vspace{-0.1cm}
The next step is to train an attack model $\mathcal{A}$ where the input would be the set of non-sensitive attributes from a target record, i.e, a $d$-dimensional vector $[x_2, ..., x_d, y]$ and the output would be a prediction for the sensitive attribute $x_1$. The adversary trains this attack model using the $DS_{\mathcal{A}}$ dataset.
The \emph{key goal} of this attack step is to learn how the target model relates the sensitive attribute with the other non-sensitive attributes including the target model's prediction label. Note that, the dataset used to train the attack model ($DS_{\mathcal{A}}$) represents a strong correlation of the sensitive attribute values with other non-sensitive ones ($[x_2, ..., x_d, y]$) since it  considers only the Case (1) records.

\subsubsection{Performing Sensitive Attribute Inference using $\mathcal{A}$} 
\label{subsec:lomia_attack}
\vspace{-0.1cm}
Once the attack model $\mathcal{A}$ is trained, the adversary can simply query $\mathcal{A}$ with the non-sensitive attributes of a target record and obtain a prediction for the sensitive attribute. It is important to note that the adversary could also query the model with the non-sensitive attributes of a record that is not in the training dataset ($DS_T$), i.e., the record is not used while training the target model.
In Section~\ref{subsec:distribution_p}, we demonstrate the effectiveness of our attacks not only in compromising the privacy of the training dataset but also their performance in breaching the distributional privacy.

%% file: Sections/V_NewAttacksPartial.tex
\vspace{-0.3cm}
\subsection{Estimating Multiple Sensitive Attributes}
\label{sec:multiple_sa}
Our LOMIA and CSMIA strategies can be easily extended to cases where the adversary aims to estimate multiple sensitive attributes of a target record. Let, $x_1, x_2$ be the sensitive attributes the adversary aims to estimate. Our strategies first perform two instances of the attacks and then stitch them together. In other words, while trying to infer $x_1$, the adversary queries the target model without setting any value for $x_2$ and vice versa~\cite{bigmlLOMIA}. 
In the case of CSMIA, we estimate the values of $x_1$ and $x_2$ independently by executing the CSMIA strategy for each of these two attributes as described in Section~\ref{subsec:cscore_attack}. 
In the case of LOMIA, we execute the LOMIA strategy independently for each of these two attributes as described in Section~\ref{subsec:lomia_dsa} and train two separate attack models to estimate the values of $x_1$ and $x_2$. The attack model to estimate $x_1$ does not take $x_2$ as an input (since the adversary does not know $x_2$) and vice versa.
Once the multiple sensitive attributes are estimated, we can also evaluate the performance of the attacks on these two attributes independently.

\vspace{-0.1cm}
\subsection{Attacks With Partial Knowledge of Target Record's Non-sensitive Attributes}
\label{sec:new_attacks_partial}
\vspace{-0.2cm}
Our  attacks proposed in this section as well as the FJRMIA~\cite{FredriksonCCS2015} strategy assume that the adversary has full knowledge of the target record's non-sensitive attributes. Although these attacks raise serious privacy concerns against a model trained on sensitive dataset, it is not clear how much risk is incurred by these model inversion attacks if the adversary has only partial access to the other non-sensitive attributes. In many cases, it may be difficult or even impossible for an adversary to obtain all of the non-sensitive attributes of a target record. Therefore, the goal of this section is to quantify the risk of MIAI attacks in the cases where all non-sensitive attributes of a target record are not known to the adversary.

Due to space constraints, in this section, we discuss only the LOMIA strategy in the case of adversary's partial knowledge of non-sensitive attributes. The discussion on how CSMIA handles this special case is described in Appendix~\ref{appn:CSMIA_partial}. 

\vspace{-0.2cm}
{
\subsubsection{LOMIA  With Partial Knowledge of Non-sensitive Attributes}
\label{subsec:LOMIA_partial}
\vspace{-0.2cm}
The attack dataset $DS_{\mathcal{A}}$ for LOMIA is obtained from Case (1) instances, i.e., the instances where only one sensitive attribute value yields the correct model prediction $y$ while all other non-sensitive attributes $x_2, ..., x_d$ remain unchanged, see Figure~\ref{fig:lomia}. Hence, the attack models in LOMIA are highly dependent on the $y$ attribute and are less dependent on other non-sensitive attributes. Therefore, even if multiple non-sensitive attributes, except the $y$ attribute, are unavailable to the attack model, the LOMIA strategy's performance does not degrade significantly. Hence, when the adversary has partial knowledge of a target record's non-sensitive attributes, the adversary can simply input the known non-sensitive attributes to the attack model and estimate the sensitive attribute. The explanation on how our attack models handle missing attributes is further discussed in Section~\ref{subsec:expres_mia_partial}.
}

%% file: Sections/VI_Evaluations.tex
\section{Evaluation}
\label{sec:expres}
\vspace{-0.2cm}
In this section, we discuss our experiment setup (i.e., datasets, machine learning models, and performance metrics) and evaluate our proposed attacks. 
To facilitate reproducibility, the links to the original datasets ($DS_T$), target models ($f$), attack model datasets ($DS_{\mathcal{A}}$), and attack models ($\mathcal{A}$) have been shared in the \emph{Availability} section. 
We will also release our codebase upon acceptance of the paper.

\subsection{Datasets}
\label{subsec:expres_data}
\vspace{-0.2cm}
\textbf{General Social Survey (GSS)~\cite{gss}}:
FJRMIA~\cite{FredriksonCCS2015} uses the \emph{General Social Survey (GSS)} dataset to demonstrate their attack effectiveness. This dataset has $51020$ records with $11$ attributes and is used to train a model that predicts how happy an individual is in his/her marriage. However, the training dataset for this model contains sensitive attribute about the individuals: e.g., responses to the question `\emph{Have you watched X-rated movies in the last year?}'. Removing the data records that do not have either the sensitive attribute or the attribute that is being predicted by the target model (i.e., happiness in marriage) results in $20314$ records that we use in our experiments. 
{
Among the $20314$ original records, $4002$ individuals answered \emph{yes} (sensitive attribute $x_1$ = \emph{yes}) to the survey question on whether they watched X-rated movies in the last year, i.e., $19.7\%$ positive class (see Table~\ref{table:data_distribution}).  
}
In order to understand if our proposed model inversion attribute inference attacks also breach the privacy of data that is not in the training dataset of the target model but is drawn from the same distribution, we split the dataset and use 75\% data to train the target models ($15235$ records in $DS_T$) and use the rest 25\% data to evaluate attacks on other data from the same distribution ($5079$ records in $DS_D$).
To ensure consistency, we evaluate other baseline attack strategies including FJRMIA~\cite{FredriksonCCS2015} on the target models trained on the $DS_T$ dataset. 
{
Among the $15235$ records in the $DS_T$ dataset, $3017$ individuals answered \emph{yes} to the question on x-rated movies, i.e., $19.8\%$ positive class (see Table~\ref{table:data_distribution}).
}

 \begin{table}[h]
  \centering
  \caption{Distribution of sensitive attributes in  datasets.}
  \resizebox{1\columnwidth}{!}{
  \begin{tabular}{ | l | l | l | l | l | l |}
    \hline
    \multirow{2}{*}{Dataset}  & Sensitive  &  Positive   & Negative   & Positive   & Positive   \\
            &     attribute  &  class label  & class label & class count   & class \%  \\ \hline
    {GSS}     & {X-movie}    & {Yes} & {No} & 4002 (3017)   & 19.7\% (19.8\%)\\
            \hline
    {Adult}  & {Marital status} & {Married} & {Single} & 21639 (16893)   & 47.8\% (47.9\%) \\
            \hline
    {{Fivethirtyeight}}  & {{Alcohol}} & {{Yes}} & {{No}} & {{266}}    & {{80.3\%}}  \\
            \hline
  \end{tabular}
  }
\label{table:data_distribution}
\end{table}

\textbf{Adult~\cite{adult}}:
This dataset, also known as \emph{Census Income} dataset, is used to predict whether an individual earns over \$50K a year. The number of instances in this dataset is $48842$ and it has $14$ attributes. We merge the `marital status' attribute into two distinct clusters, Married: \{Married-civ-spouse, Married-spouse-absent, Married-AF-spouse\} and Single: \{Divorced, Never-married, Separated, Widowed\}. We then consider this attribute (Married/Single) as the sensitive attribute that the adversary aims to learn.  After removing the data records with missing values, the final dataset consists of $45222$ records. 
Similar to the GSS dataset, we also split the Adult dataset and use $35222$ records to train the target models ($DS_T$) and use the rest $10000$ records to evaluate attacks on data from the same distribution ($DS_D$) but not in $DS_T$.
{
Among the $45222$ ($35222$) records, $21639$ ($16893$) individuals are \emph{married} (i.e., sensitive attribute $x_1$ = \emph{married}), i.e., $47.8\%$ ($47.9\%$) positive class (Table~\ref{table:data_distribution}). 
}
To ensure consistency, we evaluate all attacks in comparison against the target models trained on the $DS_T$ dataset. 
{
The `\emph{relationship}' attribute in this dataset (values: husband, wife, unmarried) is directly related to the \emph{marital status} sensitive attribute. 
Hence, for the attack setup practicality, we have removed the `\emph{relationship}' attribute from this dataset since otherwise the adversary could perform a straightforward attack: 
\texttt{if(relationship == husband || relationship == wife) then \{marital\_status = married\}
else \{marital\_status = single\}}.  
}

{
\textbf{Fivethirtyeight~\cite{538}}: 
This dataset is from a survey 
conducted by the Fivethirtyeight Datalab, also used in FJRMIA~\cite{FredriksonCCS2015}. $553$ individuals were surveyed on a variety of questions. This dataset is used to train a model that predicts how an individual would like their steak prepared. In order to evaluate the cases of estimating multi-valued and multiple sensitive attributes, we consider two sensitive attributes in this dataset: which age group an individual belongs to (multi-valued, \{18-29, 30-44, 45-60, > 60\}) and whether an individual drinks alcohol (binary, \{yes,no\}). Removing the data records missing either the sensitive attributes or the model output results in $331$ data records. We do not split this dataset further since the sample size is already small. Among $331$ individuals, $266$ answered yes to the question on drinking alcohol, i.e., $80.3\%$ positive class (Table~\ref{table:data_distribution}). The age group marginal prior distribution is $\{21.1\%, 28.1\%, 26\%, 24.8\%\}$, respectively.
}

\begin{table}[t]
  \centering
  \caption{$DS_{\mathcal{A}}$ datasets' details obtained from target models.}
  \resizebox{0.93\columnwidth}{!}{
  \begin{tabular}{ | c | c | c | c |}
    \hline
    \multirow{2}{*}{Dataset}  & Target  &  Number of    & Number of instances with correctly     \\
            &  Model    &  instances in $DS_{\mathcal{A}}$  & labeled sensitive attribute in $DS_{\mathcal{A}}$ \\ \hline
    \multirow{2}{*}{GSS}     & Decision tree    & 2387 & 1555  \\
    \cline{2 -4}            &  Deepnet  &    1011           &     564          \\ \hline

    \multirow{2}{*}{Adult}   & Decision tree    & 9263 & 7254 \\
    \cline{2 -4}            &  Deepnet  &    9960           &     7430          \\ \hline
    
    {Fivethirtyeight}   & {Decision tree}    & {49 (alcohol)} & {48} \\
    \hline
    {Fivethirtyeight}   & {Decision tree}    & {75 (age group)} & {72} \\
    \hline
  \end{tabular}
  }
\label{table:dsa}
\end{table}

\vspace{-0.2cm}
\subsection{Machine Learning Models}
\label{subsec:expres_model}
\vspace{-0.2cm}
{To ensure a fair comparison with~\cite{FredriksonCCS2015} which uses decision tree models, we first trained decision tree (DT) target models on the three datasets mentioned in Section~\ref{subsec:expres_data}. To further demonstrate the generalizability of our attacks, we also trained deep neural network (DNN) target models. 
However, we do not further use the DNN model trained on the Fivethirtyeight dataset as the model's performance is very poor due to small training set size.
The confusion matrices of all the trained models are given in Appendix (Tables~\ref{table:CM_DT_GSS},~\ref{table:CM_DNN_GSS},~\ref{table:CM_DT_Adult},~\ref{table:CM_DNN_Adult},~\ref{table:CM_DT_FTE}, and~\ref{table:CM_DNN_FTE}). Since our attack is black-box, the underlying architecture does not make any difference in our attack algorithm and so, we chose DT and DNN (the two most popular ML architecture for tabular dataset) to perform our attack on. 
}
We leverage BigML~\cite{bigml}, an ML-as-a-service system, 
and use its default configurations (1-click supervised training feature) 
to train these target models. The decision tree target models use BigML's memory tree optimization algorithm and smart pruning technique.
Each deep neural network target model has 3 hidden layers and  uses ADAM~\cite{kingma2014adam} as the optimization algorithm with a learning rate of 0.005.
The attack models of LOMIA are trained using BigML's \emph{ensemble} training algorithm with default configurations, i.e., decision forest algorithm and smart pruning technique. 
Table~\ref{table:dsa} shows the sizes of the $DS_{\mathcal{A}}$ datasets obtained from different target models along with the number of instances with correctly labeled sensitive attribute in $DS_{\mathcal{A}}$.

\begin{figure*}[t]
\centering
\subfigure[Decision tree model trained on GSS dataset]{\includegraphics[width=0.32\textwidth, height=3.2cm]
{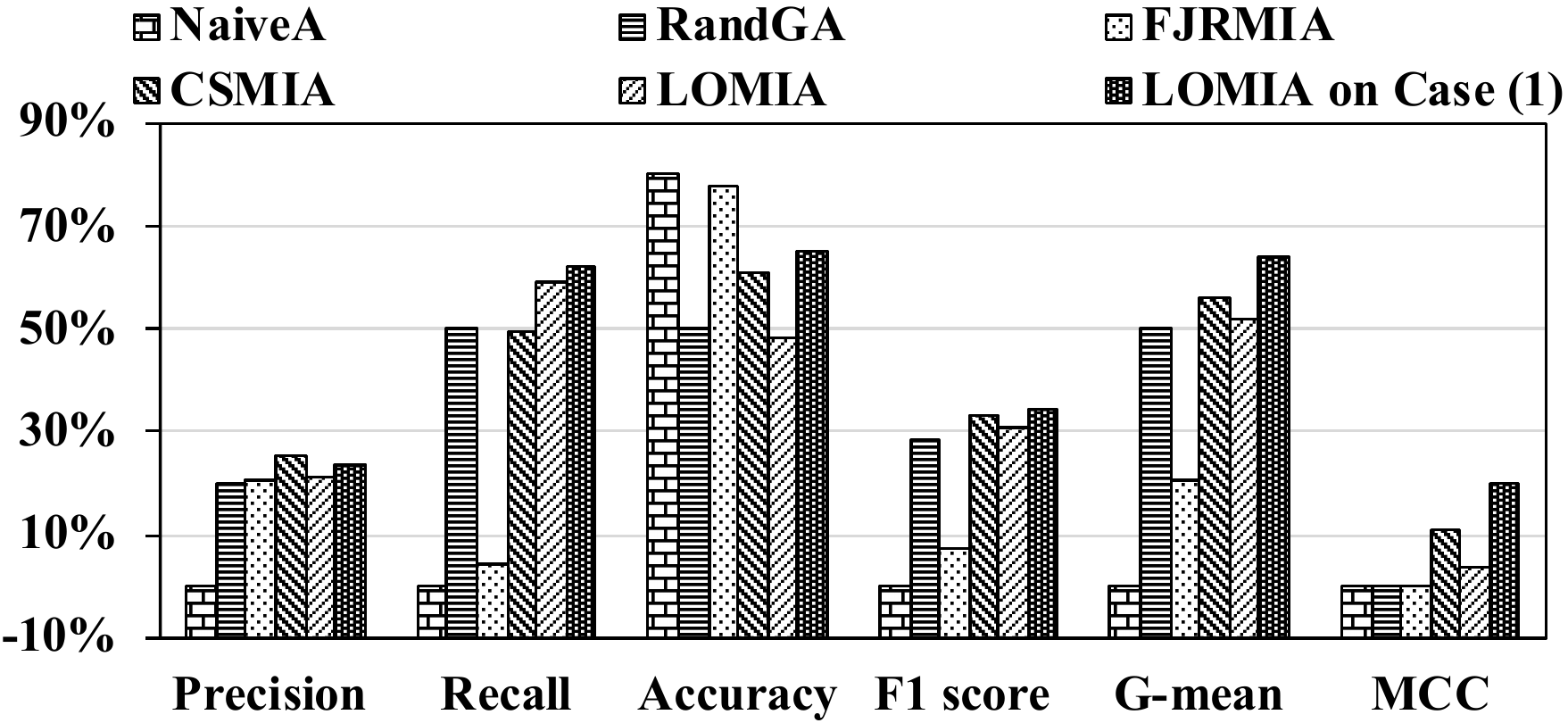}
\label{fig:GSS_RandomDT_NEW}
}
\hfill
\subfigure[Deepnet model trained on GSS dataset]{\includegraphics[width=0.32\textwidth, height=3.2cm]
{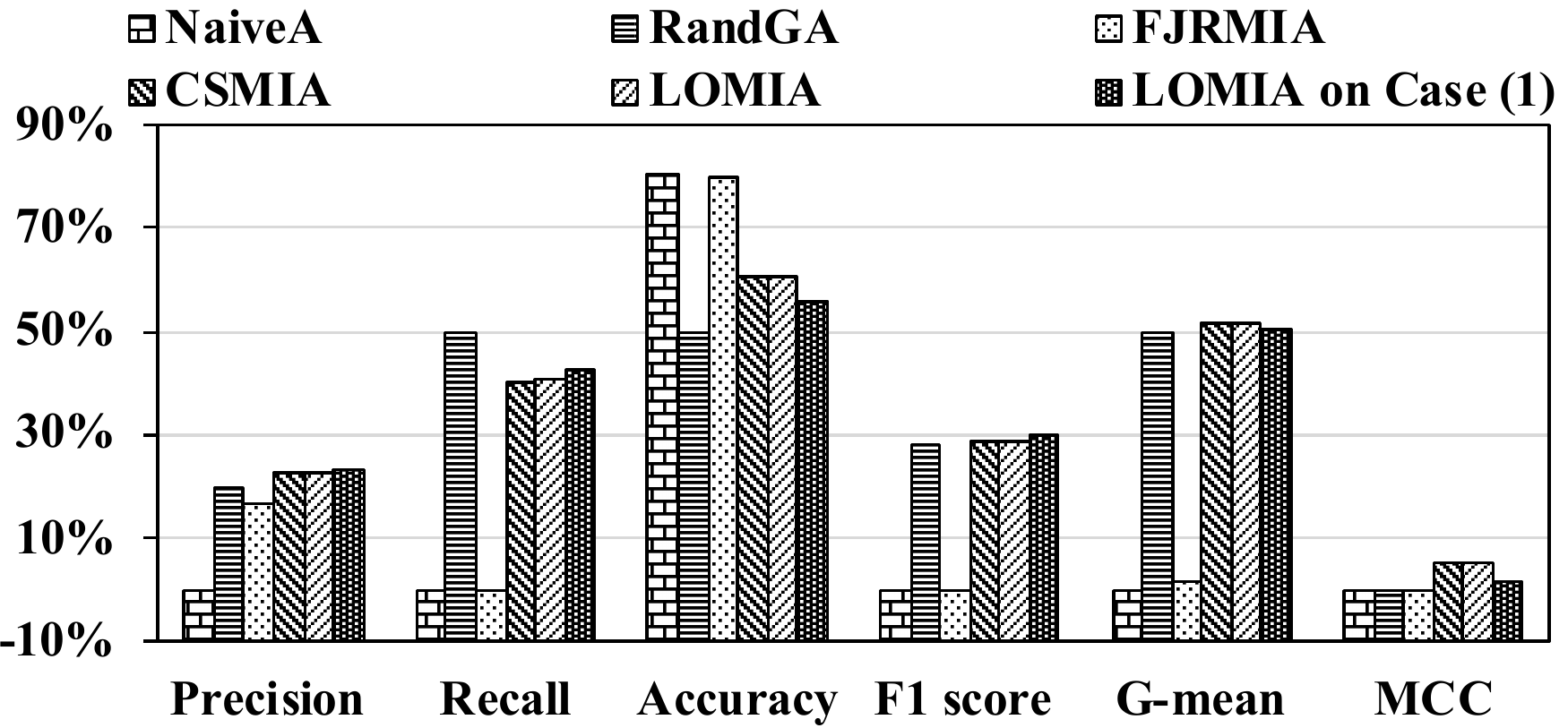}
\label{fig:GSS_RandomDNN_NEW}
}
\hfill
\subfigure[Decision tree model trained on Adult dataset]{\includegraphics[width=0.32\textwidth, height=3.2cm]
{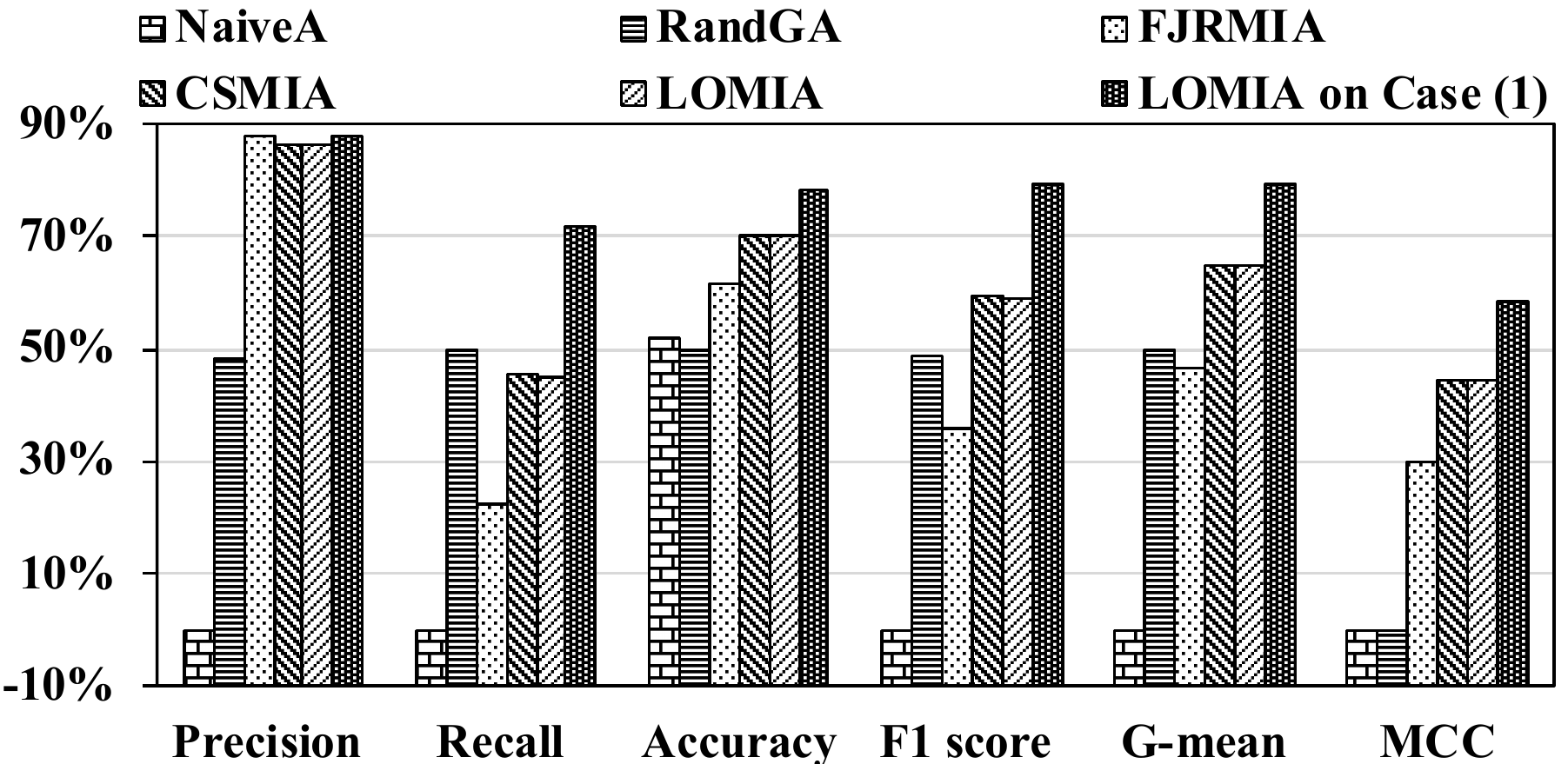}
\label{fig:Adult_RandomDT_NEW}
}
\vspace{-0.3cm}
\caption{Comparison of attacks: FJRMIA~\cite{FredriksonCCS2015}, CSMIA, and LOMIA with baseline attack strategies NaiveA and RandGA.}
\label{fig:random}
\end{figure*}

\vspace{-0.3cm}
\subsection{Attack Performance Metrics}
\label{subsec:expres_metric}
\vspace{-0.2cm}
As mentioned earlier, the \emph{accuracy} metric may fail to evaluate an attack or even misrepresent the attack performance if the dataset is unbalanced. Table~\ref{table:data_distribution} shows the distribution of sensitive attribute values in the datasets. Since the sensitive attribute in the GSS dataset is unbalanced, a naive attack always predicting the negative class would result in $\sim80\%$ accuracy, which is a misleading evaluation of attack performance. Moreover, the \emph{F1 score} alone is not a meaningful metric to evaluate the attacks since it emphasizes only on the positive class. Therefore, along with precision, recall, accuracy, and F1 score, we also use \emph{G-mean} and \emph{MCC} metrics as described in Section~\ref{sec:indepth} to evaluate our attacks on binary sensitive attributes as well as to compare their performances with that of the FJRMIA~\cite{FredriksonCCS2015} and the baseline attacks (NaiveA and RandGA). 
{
We discuss the false positive rates (FPR) of the attacks in Section~\ref{subsec:attack_stability}. In order to evaluate the proposed and existing attacks on multi-valued sensitive attributes, we compute and compare the confusion matrices of the attacks as shown in Section~\ref{subsubsec:expres_mia_fte}. 
}

{
We also evaluate the number of queries performed to the target model by the FJRMIA, CSMIA, and LOMIA strategies. For all the experiments in this section, the attacks in comparison required the same number of queries.  Section~\ref{appn:queries} in Appendix presents the details of this comparison. 
Note that, while the CSMIA extension for partial knowledge of non-sensitive attributes suffer from combinatorial complexity and make significantly more queries to the target model (Table~\ref{table:CSMIA_queries} and Appendix~\ref{appnB:CSMIA_partial}), the LOMIA strategy in the cases of partial knowledge of non-sensitive attributes does not require any extra query to the target model (see Section~\ref{subsec:LOMIA_partial}).
}

\vspace{-0.2cm}
\subsection{New Model Inversion Attacks' Results and Comparison with Baseline Attacks}
\label{subsec:expres_mia}
\vspace{-0.2cm}
In this section, we compare CSMIA and LOMIA with existing FJRMIA~\cite{FredriksonCCS2015}, and also with baseline attack strategies that do not require access to the target model, i.e., NaiveA and RandGA.  As described in Section~\ref{sec:indepth}, the goal behind comparing with NaiveA and RandGA is to understand whether releasing the black-box model really adds more advantage to the adversary to learn the sensitive attributes in the training dataset. 
We pay special attention to the Case (1) instances and analyze the LOMIA performance on them separately.

In RandGA, always predicting the positive class would result in $100\%$ recall and thus high F1 score but a G-mean of $0\%$. Therefore, for all the experiments in the following,  RandGA  predicts the positive class with a 0.5 probability, thus maximizing G-mean at $50\%$ and  ensuring a recall of $50\%$. Figures~\ref{fig:random_gss} and~\ref{fig:random_adult} in Appendix show the performance of RandGA on GSS and Adult datasets, respectively.

\vspace{-0.2cm}
\subsubsection{GSS Dataset}
\label{subsubsec:expres_mia_gss}
\vspace{-0.2cm}
Figures~\ref{fig:GSS_RandomDT_NEW} and~\ref{fig:GSS_RandomDNN_NEW} show the performances of the proposed attacks against the decision tree and deepnet target models trained on the GSS dataset, respectively, and present a comparison with FJRMIA, NaiveA, and RandGA. Table~\ref{table:gss_results_dt_dnn} in Appendix shows the details of the metrics along with the TP, TN, FP, and FN values. 
Since the sensitive attribute in this dataset has an unbalanced distribution, the NaiveA strategy, also mentioned in~\cite{FredriksonCCS2015}, predicts the sensitive attribute as \emph{no} for all the individuals and achieves an accuracy of $80.2\%$. However, the precision, recall, F1 score, G-mean, and MCC would all be $0\%$ as shown in Figures~\ref{fig:GSS_RandomDT_NEW} and~\ref{fig:GSS_RandomDNN_NEW}. 
Note that, NaiveA performance is independent of target ML model type.
As demonstrated in Figure~\ref{fig:GSS_RandomDT_NEW}, the FJRMIA~\cite{FredriksonCCS2015} achieves a very low recall and thus low F1 score. This is due to the fact that the FJRMIA~\cite{FredriksonCCS2015} relies on the marginal prior of the sensitive attribute while performing the attack. Since the sensitive attribute in the GSS dataset is unbalanced, the FJRMIA~\cite{FredriksonCCS2015} mostly predicts the negative sensitive attribute (i.e., the individual didn't watch x-rated movie, marginal prior $\sim0.8$) and rarely predicts the positive sensitive attribute (i.e., the individual watched x-rated movie, marginal prior $\sim0.2$). In contrast, our proposed CSMIA and LOMIA strategies achieve significantly high recall, F1 score, G-mean, and MCC while also improving precision. The FJRMIA~\cite{FredriksonCCS2015} performs better only in terms of accuracy. However, note that the NaiveA also achieves an accuracy of $80.2\%$, the highest among all attacks, but there is no attack efficacy (0 true positive, see Table~\ref{table:gss_results_dt_dnn}).
Our attacks also consistently outperform RandGA in terms of all metrics. We emphasize that the records that belong to Case (1) are more vulnerable to model inversion attacks. 

It is noteworthy that the LOMIA strategy performs similar to CSMIA despite having access to only the predicted labels. Unlike CSMIA, the LOMIA strategy does not have cases and uses a single attack model for all the target records. However, to better understand the contrast between the LOMIA and CSMIA strategies, we demonstrate the performance of LOMIA for the records in CSMIA cases separately (GSS case-based results in Tables~\ref{table:gss_dt_bpmpcs_details} and~\ref{table:gss_dnn_bpmpcs_details} in Appendix). 

As shown in Figure~\ref{fig:GSS_RandomDNN_NEW}, the FJRMIA~\cite{FredriksonCCS2015} strategy again achieves a high accuracy but an extremely low recall. It performs almost like NaiveA with only 1 true positive and 5 false positives (see Table~\ref{table:gss_results_dt_dnn}). The RandGA strategy has the same results as Figure~\ref{fig:GSS_RandomDT_NEW} since this strategy is independent of the target model (similar to NaiveA). Our attacks' performances against this model are not significantly better than RandGA, even the LOMIA results on Case (1) are not significant. Therefore, it may seem that according to the overall performance, the deep neural network model trained on the GSS dataset may not be vulnerable to model inversion attacks since the RandGA attack even without access to the model may achieve comparable performances. However, it is very important to note that the RandGA strategy predicts the sensitive attribute randomly whereas the model inversion attacks rely on the outputs of a model that is trained on the dataset containing the actual sensitive attributes. Even if the overall performance of a model inversion attack on the entire dataset does not seem to be a threat, some specific groups of records (e.g., individuals grouped by race, gender) in the dataset could still be vulnerable. We discuss such discrimination in performances of model inversion attacks later in Section~\ref{subsec:expres_target}.

\vspace{-0.2cm}
\subsubsection{Adult Dataset}
\label{subsubsec:expres_mia_adult}
\vspace{-0.2cm}
Figure~\ref{fig:Adult_RandomDT_NEW} shows the performances of the attacks against the decision tree target model trained on the Adult dataset. The results for deepnet target model are very similar to that of decision tree (see Figure~\ref{fig:Adult_RandomDNN_NEW} in Appendix).
Table~\ref{table:adult_results_dt_dnn} in Appendix shows the details along with the TP, TN, FP, and FN values. 
Since the sensitive attribute is more balanced in this dataset, the NaiveA strategy has an accuracy of only $52.1\%$, and the other metrics are at $0\%$.
FJRMIA~\cite{FredriksonCCS2015} results in a precision comparable to our attacks but achieves much less in terms of the other metrics. Our attacks also significantly outperform RandGA in terms of all metrics except the recall. 

Tables~\ref{table:adult_dt_bpmpcs_details} and~\ref{table:adult_dnn_bpmpcs_details} 
in Appendix show the contrast between CSMIA and LOMIA in details.
Observing the results of the proposed attacks and also the performance against Case (1) instances, we conclude that \emph{releasing the models trained on the Adult dataset} would add significant advantage to the adversary in terms of learning the `marital status' sensitive attribute. This is because all our proposed attacks 
that query the models for sensitive attribute inference 
perform significantly better when compared to the NaiveA and RandGA adversary that do not need any access to the model.

Overall, the attacks against the target models trained on the Adult dataset demonstrate more effectiveness than that of against the target models trained on the GSS dataset. Therefore, we investigated if the correlations between the sensitive attributes and the corresponding target models trained on these datasets (in other words, the importance of the sensitive attributes in the target models) differ significantly. However, according to our observation, this is not the case. For instance, the importance of the `x-rated-movie' and `marital-status' sensitive attributes in their corresponding decision tree target models are $7.3\%$ and $9.6\%$, respectively. Figure~\ref{fig:importance} in Appendix shows the importance of all attributes in these models.

\begin{table*}[t]
    \caption{Attacks against DT target model trained on FiveThirtyEight dataset to infer `age' sensitive attribute, attack confusion matrices of (a) FJRMIA, (b) CSMIA, (c) LOMIA, and (d) LOMIA (Case 1).}
    \label{table:CM_DT_FTE_2}
    \vspace{-0.3cm}
    \begin{minipage}{.49\linewidth}
      \subfigure[]{
      \centering
          \resizebox{\textwidth}{!}{
            \begin{tabular}{|l|*{6}{c|}}\hline
                \backslashbox{Actual}{Predicted}
                &18-29&30-44&45-60&>60& Total & Recall \\\hline
                18-29 & 0 & 70 & 0 & 0 & 70 & 0\% \\\hline 
                30-44  & 0 & 93 & 0 & 0 & 93 & 100\% \\\hline 
                45-60& 0 & 86 & 0 & 0 & 86 & 0\% \\\hline 
                >60& 0 & 82 & 0 & 0 & 82 & 0\% \\\hline 
               Total         & 0 & 331 & 0 & 0 & 331 & Avg. rec. 25\% \\\hline 
                Precision     & 0\% & 28.1\% & 0\% & 0\% & Avg. prec. 7.02\% & Accuracy 28.1\% \\\hline 
            \end{tabular}
            \label{table:CM_DT_FTE_2_FJR}
            }
            }
    \end{minipage}%
    \hfill
    \begin{minipage}{.49\linewidth}
      \subfigure[]{
      \centering
          \resizebox{\textwidth}{!}{
            \begin{tabular}{|l|*{6}{c|}}\hline
                \backslashbox{Actual}{Predicted}
                &18-29&30-44&45-60&>60& Total & Recall \\\hline
                18-29 & 40 & 9 & 8 & 13 & 70 & 57.14\% \\\hline 
                30-44  & 13 & 49 & 12 & 19 & 93 & 52.69\% \\\hline 
                45-60& 15 & 17 & 36 & 18 & 86 & 41.86\% \\\hline 
                >60& 11 & 19 & 21 & 31 & 82 & 37.8\% \\\hline 
               Total         & 79 & 94 & 77 & 81 & 331 & Avg. rec. 47.37\% \\\hline 
                Precision     & 50.63\% & 52.13\% & 46.75\% & 38.27\% & Avg. prec. 46.95\% & Accuracy 47.13\% \\\hline 
            \end{tabular}
            \label{table:CM_DT_FTE_2_CS}
            }
            }
    \end{minipage}%
    
    \begin{minipage}{.49\linewidth}
      \subfigure[]{
      \centering
          \resizebox{\textwidth}{!}{
            \begin{tabular}{|l|*{6}{c|}}\hline
                \backslashbox{Actual}{Predicted}
                &18-29&30-44&45-60&>60& Total & Recall \\\hline
                18-29 & 41 & 20 & 9 & 0 & 70 & 58.57\% \\\hline 
                30-44  & 21 & 50 & 18 & 4 & 93 & 53.76\% \\\hline 
                45-60& 28 & 24 & 32 & 2 & 86 & 37.21\% \\\hline 
                >60& 30 & 30 & 12 & 10 & 82 & 12.2\% \\\hline 
               Total         & 120 & 124 & 71 & 16 & 331 & Avg. rec. 40.43\% \\\hline 
                Precision     & 34.17\% & 40.32\% & 45.07\% & 62.5\% & Avg. prec. 45.51\% & Accuracy 40.18\% \\\hline 
            \end{tabular}
            \label{table:CM_DT_FTE_2_LO}
            }
            }
    \end{minipage}%
    \hfill
    \begin{minipage}{.49\linewidth}
      \subfigure[]{
      \centering
          \resizebox{\textwidth}{!}{
            \begin{tabular}{|l|*{6}{c|}}\hline
                \backslashbox{Actual}{Predicted}
                &18-29&30-44&45-60&>60& Total & Recall \\\hline
                18-29 & 21 & 0 & 0 & 0 & 21 & 100\% \\\hline 
                30-44  & 0 & 23 & 0 & 0 & 23 & 100\% \\\hline 
                45-60& 1 & 0 & 19 & 1 & 21 & 90.48\% \\\hline 
                >60& 1 & 0 & 0 & 9 & 10 & 90\% \\\hline 
               Total         & 23 & 23 & 19 & 10 & 75 & Avg. rec. 95.12\% \\\hline 
                Precision     & 91.3\% & 100\% & 100\% & 90\% & Avg. prec. 95.33\% & Accuracy 96\% \\\hline 
            \end{tabular}
            \label{table:CM_DT_FTE_2_LO1}
            }
            }
    \end{minipage}%
    \vspace{-0.5cm}
\end{table*}

\vspace{-0.2cm}
\subsubsection{FiveThirtyEight Dataset}
\label{subsubsec:expres_mia_fte} 
\vspace{-0.2cm}
In this section, we perform two sets of attack experiments against the DT target model trained on the FiveThirtyEight dataset: 
(i) inferring multi-valued sensitive attribute \emph{age group}, when all other non-sensitive attributes are known to the adversary, and
(ii) inferring both \emph{alcohol} and \emph{age group}, i.e., the case of estimating multiple sensitive attributes.

\begin{enumerate}[label=\textbf{(\roman*)},  leftmargin=0pt, itemindent=15pt]
\item  {\textbf{Estimating Multi-valued Sensitive Attributes}}

\noindent {Tables~\ref{table:CM_DT_FTE_2} (a), (b), and (c) show the performances of the FJRMIA, CSMIA, and LOMIA strategies, respectively, in terms of estimating a multi-valued sensitive attribute, i.e., \emph{age} in the FiveThirtyEight dataset. FJRMIA~\cite{FredriksonCCS2015} predicts the age group $30-44$ for all the target records (i.e., it boils down to NaiveA, age group $30-44$ has the highest marginal prior among all, $28.1\%$). Also, RandGA strategy would achieve a maximum accuracy of $25\%$ in estimating this multi-valued sensitive attribute (not shown in tables). In contrast, our proposed CSMIA and LOMIA strategies achieve significantly better results. The results in Table~\ref{table:CM_DT_FTE_2} (d) show the performance of LOMIA on Case (1) instances which has an accuracy of $96\%$. Hence, we emphasize that the records in Case (1) are significantly more vulnerable to model inversion attacks. 
}

\vspace{0.2cm}
\item  {\textbf{Estimating Multiple Sensitive Attributes}}

\noindent  {In this attack setting, the adversary estimates both the age group and alcohol sensitive attributes of a target individual.
The attack results for estimating the multi-valued age group attribute in this case are similar to that of Table~\ref{table:CM_DT_FTE_2}. Due to space constrains, we demonstrate the performances of the FJRMIA, CSMIA, and LOMIA strategies in terms of estimating the age group attribute in Tables~\ref{table:CM_DT_FTE_3_FJRMIA_app},~\ref{table:CM_DT_FTE_3_CSMIA_app}, and~\ref{table:CM_DT_FTE_3_LOMIA_app} in Appendix, respectively. The attack results for estimating the binary attribute alcohol  are given in Table~\ref{table:fte_exp3_results_dt}.
}

\end{enumerate}

\begin{figure}[h]
\centering
\includegraphics[width=0.99\columnwidth, height=3.2cm]
{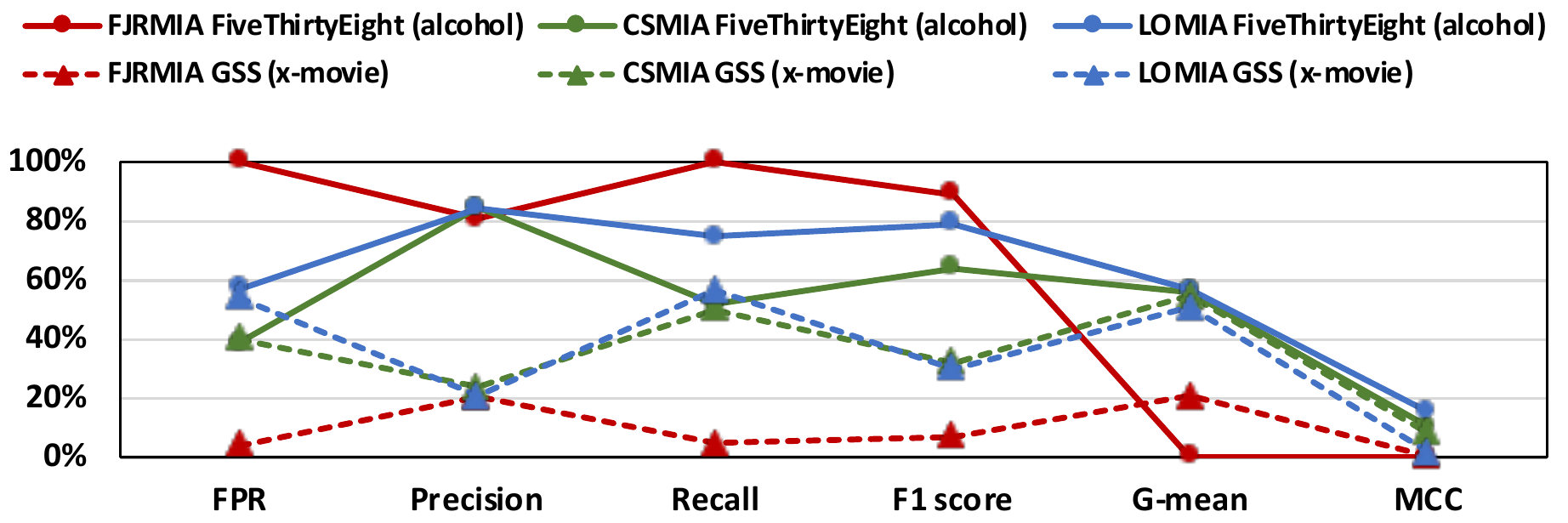}
\vspace{-0.1cm}
\caption{{Comparison among different attack strategies in terms of FPR and other metrics}}
\label{fig:stability}
\end{figure}

\vspace{-0.4cm}
{
\subsection{False Positive Rates and Attack Stability}
\label{subsec:attack_stability} 
\vspace{-0.1cm}
In order to demonstrate the false positive rate (FPR) comparison between our proposed attacks and the existing FJRMIA~\cite{FredriksonCCS2015} strategy, we perform experiments with two scenarios: (1) estimating the `alcohol' sensitive attribute in the FiveThirtyEight dataset which has $80.3\%$ positive class marginal prior (i.e., alcohol=yes), and (2) estimating the `x-movie' sensitive attribute in the GSS dataset which has only $19.8\%$ positive class marginal prior (i.e., x-movie=yes).
Figure~\ref{fig:stability} shows the comparison among FJRMIA, CSMIA, and LOMIA in terms of FPR and other metrics. The solid lines represent the attack performances of estimating alcohol in the FiveThirtyEight dataset whereas the dashed lines represent the attack performances of estimating x-movie in the GSS dataset. Since FJRMIA is heavily dependent on the marginal priors of the sensitive attributes, it achieves extreme FPRs in these two scenarios: $100\%$ FPR in estimating alcohol and $4.17\%$ FPR is estimating x-movie. 
In contrast, our proposed attacks are more stable and their superior performance in both scenarios are evident by the G-mean and MCC metrics in Figure~\ref{fig:stability}. 
The comparison of these attacks' FPRs for Adult dataset where the sensitive attribute is more balanced is given in Table~\ref{table:adult_results_dt_dnn}. The FPRs of our proposed attacks are comparable to that of FJRMIA ($\sim6\%$ vs. $\sim3\%$). However, our attacks outperform FJRMIA in terms of other metrics as shown in Figures~\ref{fig:Adult_RandomDT_NEW} and~\ref{fig:Adult_RandomDNN_NEW}.
Note that, lower FPR may not always indicate better attack, e.g., NaiveA has an FPR of $0\%$ but the attack has no efficacy.
}

\begin{figure*}[t]
\centering
\subfigure[]
{\includegraphics[width=0.25\textwidth, height=3cm]
{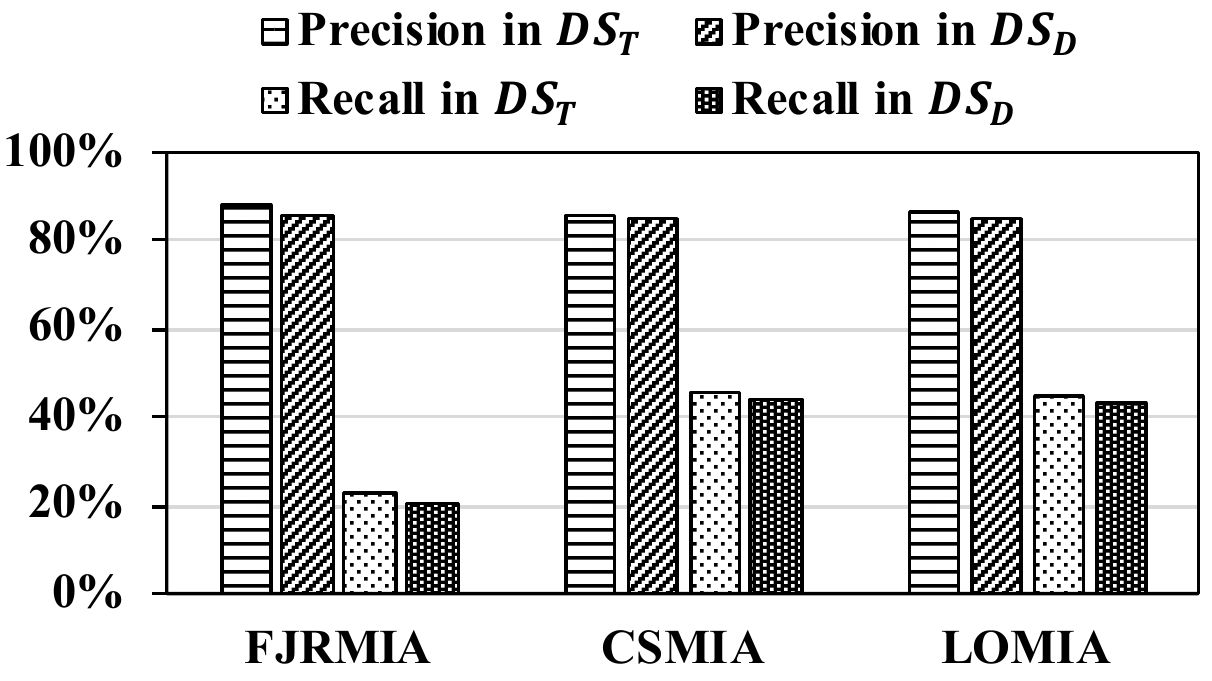}
\label{fig:dis_privacy}
}
\hfill
\subfigure[]
{\includegraphics[width=0.72\textwidth, height=2.5cm]
{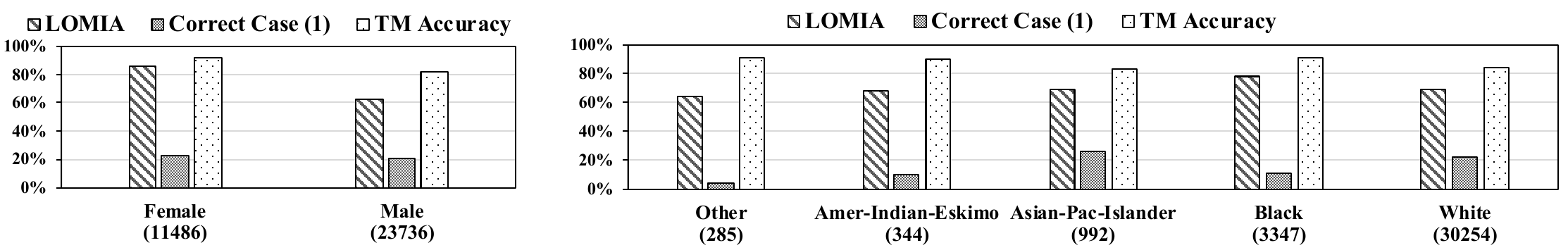}
\label{fig:disparate_gender_race}
}
\vspace{-0.3cm}
\caption{(a) Privacy leakage for $DS_T$ and $DS_D$, (b) disparate vulnerability of LOMIA for different gender and race groups.}
\vspace{-0.3cm}
\label{fig:mixed}
\end{figure*}

\vspace{-0.2cm}
\subsection{Distributional Privacy Leakage}
\label{subsec:distribution_p}
\vspace{-0.2cm}
In order to investigate if our MIAI attacks also breach the privacy of data that is not in the training dataset of the target model but is drawn from the same distribution, we evaluate our attacks on the corresponding $DS_D$ datasets as described in Section~\ref{subsec:expres_data}. Figure~\ref{fig:dis_privacy} compares the performance of our attacks as well as the performance of FJRMIA on the decision tree model trained on Adult dataset. Our observation shows that our attacks are equally effective against the records in the training dataset ($DS_T$) and the records outside of the training dataset but drawn from the same distribution ($DS_D$). 
{
We observe similar trends in the proposed attacks against other target models as shown in Figure~\ref{fig:distributional_p} in Appendix.
}

\vspace{-0.2cm}
\subsection{Disparate Vulnerability of MIAI Attacks}
\label{subsec:expres_target}
\vspace{-0.2cm}
In this section, we further investigate the vulnerability of model inversion attacks by analyzing the attack performances on different groups in the dataset. If a particular group in a dataset is more vulnerable to these attacks than others, it raises serious privacy concerns for that particular group. 

Figure~\ref{fig:disparate_gender_race} shows the contrast in the  performances of LOMIA against different gender and race populations. The attack is performed against the deepnet model trained on Adult dataset. The x-axis represents gender/race identities along with the number of records in the training dataset that belong to the particular subgroups. For instance, the numbers of female and male individuals in the Adult dataset are $11,486$ and $23,736$, respectively. 
{
According to our observation, LOMIA could predict the correct marital status for $85.9\%$ of the female population whereas it could predict the correct marital status for only $62.4\%$ of the male population. 
LOMIA also shows disparate attack performance against different race groups, and is most successful against the Black race subgroup with $78.2\%$ accuracy.
Since the attack model of LOMIA is trained on $DS_{\mathcal{A}}$ dataset obtained from the Case (1) instances, we investigated what percentage of records of each of the female and male subgroups are labeled with correct sensitive attributes in $DS_{\mathcal{A}}$ dataset and if that has any impact on such disparate vulnerability.
However, we observe that around a similar percentage ($\sim21\%$) of both female and male records, i.e., $2593$ and $4837$, respectively, are labeled with the correct sensitive attribute (single/married) in the $DS_{\mathcal{A}}$ dataset, which is shown using Correct Case (1) bar in  Figure~\ref{fig:disparate_gender_race}.
We also investigated if accuracy of target model for different subgroups plays a role in disparate vulnerability, shown using TM Accuracy bar in Figure~\ref{fig:disparate_gender_race}. We observe that  target model is $92.4\%$ accurate for the female population and only $81.4\%$ accurate for the male population in predicting their income, which correlates with the disparate vulnerability. However, we have not observed this correlation consistently, e.g., in the case of disparate vulnerability for race subgroups.
}
LOMIA shows disparate vulnerability against other subgroups, such as {religions (DT model trained on GSS dataset)} and occupations (DNN model trained on Adult dataset). The results are demonstrated in Appendix (see Figures~\ref{fig:disp_gss_rel} and~\ref{fig:disp_adult_occ} in Appendix~\ref{appnC}, respectively).
{
Note that, we have observed disparate-vulnerability across all datasets and models but reported the most interesting results only.
}

The performance of an adversary with RandGA strategy would not differ significantly for these different groups because of their random prediction. Due to the differences in the underlying distributions of the married individuals in these groups, the RandGA strategy would only show slightly different performance in terms of precision and thus in the F1 score. 
While our findings here show only a few instances of such disparity in the model inversion attack performances on different groups, this is a potentially serious issue and needs to be further investigated.  Otherwise, while it may seem that the attack performance on the overall dataset is not a significant threat, some specific groups in the dataset could still remain significantly vulnerable to MIAI attacks.

\begin{figure}[t]
\centering
\includegraphics[width=1\columnwidth, height=3cm]
{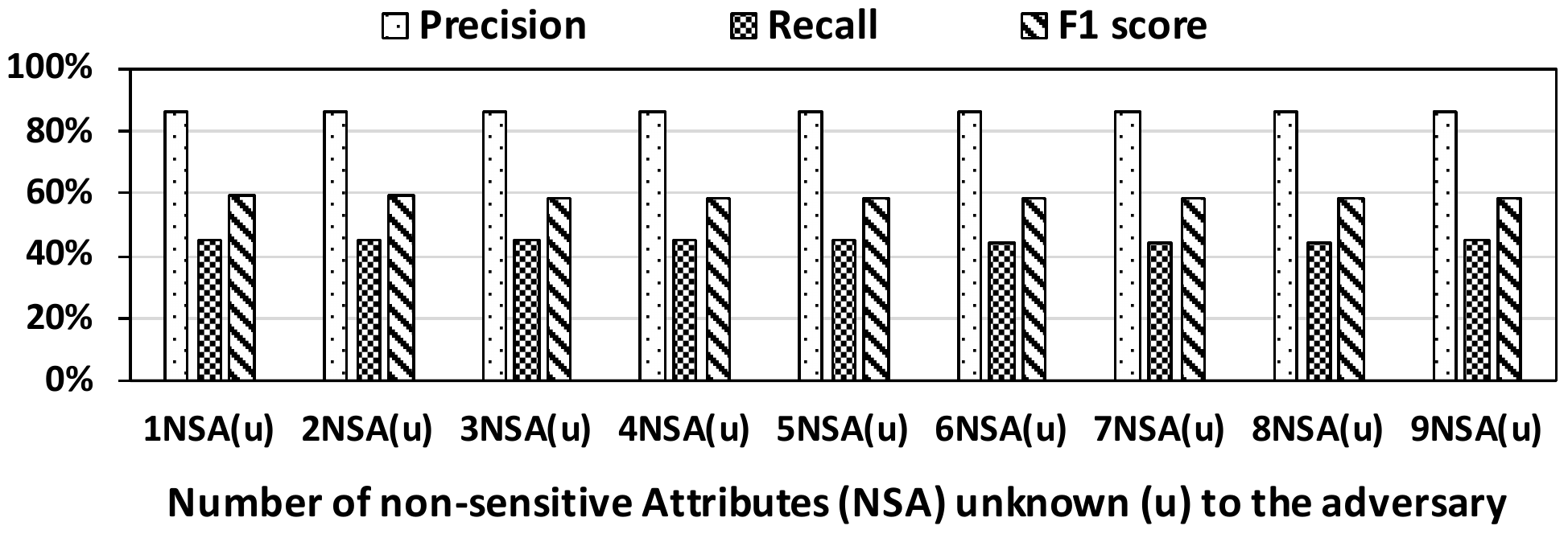}
\vspace{-0.5cm}
\caption{LOMIA performance against the decision tree model trained on Adult dataset when 1-9 non-sensitive attributes (NSA) are unknown (u) to the adversary.
}
\label{fig:lomia_adult_dt_partial}
\end{figure}

\vspace{-0.2cm}
\subsection{Attack Results With Partial Knowledge of Target Record's Non-sensitive Attributes}
\label{subsec:expres_mia_partial}
\vspace{-0.1cm}
{
With partial knowledge of target record's non-sensitive attributes, our LOMIA ensemble attack models handle the missing attributes using \emph{last prediction strategy}~\cite{bigmlLOMIA}. With this strategy, the prediction is computed by descending the branches of the tree according to the available input attributes. When there is a question regarding the missing attribute, the process stops and the prediction of the last node is returned.
}

{
Figure~\ref{fig:lomia_adult_dt_partial} shows the performance details of LOMIA against the decision tree model trained on Adult dataset when 1-9 non-sensitive attributes (NSA) increasingly become unknown (u) to the adversary in the following order: work-class, sex, race, fnlwgt, occupation, education, hours-per-week, capital-gain, and capital-loss. This order reflects the importance of the Adult dataset attributes in the LOMIA attack model trained against the decision tree target model (see Figure~\ref{fig:importance_adult} (a)). Since the `income' attribute occupies $90.4\%$ importance in the LOMIA attack model, unavailability of  9 other non-sensitive attributes does not degrade the performance of LOMIA. We have observed similar LOMIA results against other target models. Figures~\ref{fig:importance_adult} (b),~\ref{fig:importance_gss} (a),~\ref{fig:importance_gss} (b) in Appendix show the importance of the dataset attributes in the LOMIA attack models. Figures~\ref{fig:lomia_adult_dnn_partial},~\ref{fig:lomia_gss_dt_partial}, and~\ref{fig:lomia_gss_dnn_partial} in Appendix show the performance details of LOMIA, respectively.
}

\emph{These results not only show an increased vulnerability of  model inversion attacks but also escalate the practicability of such attacks in the real world where the adversary may not know all other attributes of a target record.}
{
Due to space constraints, the performance details of the CSMIA partial knowledge attack have been discussed in Appendix~\ref{appnB:CSMIA_partial}.
}

\begin{table}[t]
  \centering
  \caption{Attack performance against the decision tree target model trained on Adult dataset.}
  \resizebox{.99\columnwidth}{!}{
  \begin{tabular}{ | l | l | l | l | l | l | l | l | l | l |}
    \hline
    Target model & \multirow{2}{*}{Attack Strategy} &  \multirow{2}{*}{TP} & \multirow{2}{*}{TN} & \multirow{2}{*}{FP} & \multirow{2}{*}{FN} & \multirow{2}{*}{Precision} & \multirow{2}{*}{Recall} & \multirow{2}{*}{Accuracy} & \multirow{2}{*}{F1 score}  \\
    class label & & & & & & & & &  \\ \hline 
    
    \multirow{3}{*} {<=50K} & FJRMIA~\cite{FredriksonCCS2015} & $13$ & $17108$ & $13$ & $9315$ & $50\%$ & $0.14\%$ & $64.73\%$ & $0.28\%$  \\ 
    \cline{2-10} & CSMIA & $127$ & $17018$ & $103$ & $9201$ & $55.22\%$ & $1.36\%$ & $64.82\%$ & $2.66\%$ \\ 
    \cline{2-10} & LOMIA & $26$ & $17085$ & $36$ & $9302$ & $41.94\%$ & $0.28\%$ & $64.69\%$ & $0.55\%$  \\ \hline 
    
    \multirow{3}{*} {>50K} & FJRMIA~\cite{FredriksonCCS2015} & $3775$ & $710$ & $498$ & $3790$ & $88.34\%$ & $49.9\%$ & $51.12\%$ & $63.78\%$  \\
    \cline{2-10} & CSMIA & $7537$ & $67$ & $1141$ & $28$ & $86.85\%$ & $99.63\%$ & $86.68\%$ & $92.8\%$ \\ 
    \cline{2-10} & LOMIA  & $7548$ & $47$ & $1161$ & $17$ & $86.67\%$ & $99.78\%$ & $86.57\%$ & $92.76\%$  \\ \hline

  \end{tabular}
  }
\label{table:class_label}
\end{table}

\subsection{Attacks' Efficacy on Different Class Labels of Target Model}
\label{subsec:model_vs_attack}
In this section, we aim to understand the efficacy of model inversion attacks for different class labels of the target model  and focus on the decision tree model trained on Adult dataset.

Table~\ref{table:class_label} shows a comparison among FJRMIA~\cite{FredriksonCCS2015}, CSMIA, and LOMIA performances for different class labels of the target model. Note that, the attack performances are significantly different for the two class labels, e.g., the recall values of identifying `married' individuals in class <=50K are significantly low when compared to the recall values of identifying `married' individuals in class >50K. The precision values also demonstrate disparate attack performances on these two target model class labels.

\subsection{Discussion and Limitation}
\label{subsec:discussion}
\vspace{-0.2cm}
To our knowledge, ours is the first work that studies MIAI attacks  in such details on \emph{tabular} datasets which is the most common data type used in real-world machine learning~\cite{luo2020network}. 
We discuss some of our notable findings in the following:

\noindent {\textbf{TIR vs. MIAI}:} As mentioned in Section~\ref{sec:intro}, the TIR attacks have strong correlations with the model's predictive power. This is because highly predictive models are able to establish a strong correlation between features and labels, and this is the property that an adversary exploits to mount the TIR attacks~\cite{Zhang_2020_CVPR}. However, we argue that such is \emph{not} the case for MIAI attacks.  
Table~\ref{table:CM_DT_Adult} in Appendix shows the confusion matrix for the decision tree model trained on Adult dataset. From the matrix, it is evident that the target model's performance (both precision and recall) is better for class label $<=50K$ than that of for class label $>50K$. If the root causes of MIAI attacks were similar to that of TIR attacks, the attacks would be more effective against the records of class label $<=50K$. On the contrary, in Section~\ref{subsec:model_vs_attack}, we demonstrate that the MIAI attacks (both existing and proposed) perform better against the records of class label $>50K$.

\noindent {\textbf{Importance of sensitive attribute in target model}:} As discussed in Section~\ref{subsubsec:expres_mia_adult}, the importances of sensitive attributes in the corresponding target models trained on GSS and Adult datasets do not differ significantly whereas the proposed MIAI attacks against target models trained on Adult dataset are significantly more effective than that of against the target models trained on GSS dataset. This indicates that only controlling the \emph{importance} of the sensitive attributes in the target model may not be always sufficient to reduce the risk of model inversion attacks. 
{
We identify the difference in the distribution of sensitive attributes in these datasets (Adult dataset $47.9\%$ positive class vs. GSS dataset $19.8\%$ positive class) as a factor that has contributed to this attack performance difference. We leave investigating this and other factors to future work.
}

\noindent {\textbf{Disparate vulnerability}: We have investigated correct Case (1) percentage and target model accuracy for different subgroups as possible factors behind disparate vulnerability.} It is evident that further investigation is required to better understand the disparate impact on different groups of records which is a serious threat of model inversion attacks.

\noindent {\textbf{Distributional privacy breach}: Existing research~\cite{PharmaUSENIX2014, Zhang_2020_CVPR} shows that differential privacy (DP)-based defense mechanisms against model inversion attacks suffer from significant loss of model utility.} Moreover, DP mechanisms provide privacy guarantees to only the training data records. In contrast, our experiments show that model inversion attacks not only breach the privacy of sensitive training dataset but also leaks distributional privacy. {Therefore, the effectiveness of DP mechanisms against model inversion attacks needs further investigation.}

\noindent {\textbf{Limitations}:
Attribute inference attack is not a realistic threat when a dataset has a lot of attributes, since model prediction is likely to depend very little on each individual attribute.  Therefore, in this paper, we  study the MIAI attacks only on datasets with fewer attributes. 
}

%% file: Sections/VII_RelatedWork.tex
\section{Related Work}
\label{sec:related}
\vspace{-0.3cm}
In~\cite{PharmaUSENIX2014}, Fredrikson et al. introduced the concept of model inversion attacks and applied their attack to linear regression models. In~\cite{FredriksonCCS2015}, Fredrikson et al. extended their attack so that it could also be applicable to non-linear models, such as decision trees. The later work presents two types of applications of the model inversion attack. The first one assumes an adversary who has access to a model (for querying) and aims to learn the sensitive attributes in the dataset that has been used to train that model (also known as attribute inference attack). In the second setting, the adversary aims to reconstruct instances similar to ones in the training dataset using gradient descent. Particularly, their attack generates images similar to faces used to train a facial recognition model. As mentioned earlier, we focus on the first one, i.e., attribute inference attack. Subsequently, Wu et al.~\cite{wu2016methodology} presented a methodology to formalize the model inversion attack.

A number of attribute inference attacks have been shown to be effective in different domains, such as social media~\cite{GongTOPS2018, AttriInferWWW2017, GongUSENIX2016, GongTIST2014, KosinskiPNAS2013, ChaabaneNDSS2012, SNPrivacyWWW2009} and recommender systems~\cite{weinsberg2012blurme, wu2020joint}. In the case of social media, the adversary infers the private attributes of a user (e.g., gender, political views, locations visited) by leveraging the knowledge of other attributes of that same user that are shared publicly (e.g., list of pages liked by the user, etc). The adversary first trains a machine learning classifier that takes as input the public attributes and then outputs the private attributes.
However, in order to build such a classifier, these attacks~\cite{GongTOPS2018, AttriInferWWW2017, GongUSENIX2016, GongTIST2014, KosinskiPNAS2013, ChaabaneNDSS2012, SNPrivacyWWW2009} have to rely on social media users who also make their private attributes public. Therefore, the adversary's machine learning classifier can be built only in those scenarios where it can collect the private-public attribute pairs of real users. Also, for the attacks shown in the recommender systems~\cite{weinsberg2012blurme}, at first, the adversary has to collect data of the users who also share their private attributes (e.g., gender) publicly along with their public rating scores (e.g., movie ratings). In contrast to the adversaries assumed in these attacks~\cite{GongTOPS2018, AttriInferWWW2017, GongUSENIX2016, GongTIST2014, KosinskiPNAS2013, ChaabaneNDSS2012, SNPrivacyWWW2009,weinsberg2012blurme}, the adversaries assumed in our attacks are \emph{not} assumed to be able to obtain a dataset from the same population the $DS_T$ dataset has been obtained from. This is because in many scenarios such an assumption (adversary having access to a similar dataset) may not be valid. Therefore, while designing our attacks, it has been part of our goal to incorporate these practical scenarios in our attack surface so that our proposed attacks could be applied more widely.

Shokri et al.~\cite{shokri2019privacy} investigate whether transparency of machine learning models conflicts with privacy and demonstrate that record-based explanations of machine learning models can be effectively exploited by an adversary to reconstruct the training dataset. In their setting, the adversary can generate unlimited transparency queries and for each query, the adversary is assumed to get in return some of the original training dataset records (that are related to the queries) as part of the transparency report. 
He et al.~\cite{HeACSAC2019} devise a new set of model inversion attacks against collaborative inference where a deep neural network and the corresponding inference task are distributed among different participants. The adversary, as a malicious participant, can accurately recover an arbitrary input fed into the model, even if it has no access to other participants' data or computations, or to prediction APIs to query the model. 

Most of the work mentioned above assume that the attributes of a target individual, except the sensitive attribute, are known to the adversary. Hidano et al.~\cite{HidanoPST2017} proposed a method to infer the sensitive attributes without the knowledge of non-sensitive attributes. However, they consider an online machine learning model and assume that the adversary has the capability to poison the model with malicious training data. In contrast, our model inversion attack with partial knowledge of target individual's non-sensitive attributes does not require poisoning and performs similar to scenarios where the adversary has full knowledge of target individual's non-sensitive attributes.

Zhang et al.~\cite{Zhang_2020_CVPR} present a generative model-inversion attack to invert deep neural networks. They demonstrate the effectiveness of their attack by reconstructing face images from a state-of-the-art face recognition classifier. They also prove that a model’s predictive power and its vulnerability to inversion attacks are closely related, i.e., highly predictive models are more vulnerable to inversion attacks.
Aïvodji et al.~\cite{avodji2019gamin} introduce a new black-box model inversion attack framework,
GAMIN (Generative Adversarial Model INversion), based on the continuous training of
a surrogate model for the target model and  evaluate their attacks on
convolutional neural networks. In~\cite{YangCCS2019}, Yang et al. train a second neural network that acts as the inverse of the target model while assuming partial knowledge about the target model's training data. The objective of the works mentioned above is typical instance reconstruction (TIR), i.e., similar to the second attack mentioned in~\cite{FredriksonCCS2015}.

%% file: Sections/VIII_Conclusions.tex
\section{Conclusion and Future Work}
\vspace{-0.3cm}
In this paper, we demonstrate two new black-box model inversion attribute inference (MIAI) attacks: (1) confidence score-based attack (CSMIA) and (2) label-only attack (LOMIA). 
The CSMIA strategy assumes that the adversary has access to the target model's confidence scores whereas the LOMIA strategy assumes the adversary's access to the label predictions only. Despite access to only the labels, our label-only attack performs on par with the proposed confidence score-based MIAI attack. 
Along with accuracy and F1 score, we propose to use the G-mean and Matthews correlation coefficient (MCC) metrics in order to ensure effective evaluation of our attacks as well as the state-of-the-art attacks. We perform an extensive evaluation of our attacks using two types of machine learning models, decision tree and deep neural network, that are trained  {with three real datasets~\cite{gss, adult, 538}}.  Our evaluation results show that the proposed attacks significantly outperform the existing ones. Moreover, we empirically show that model inversion attacks have disparate vulnerability property and consequently, a particular subset of the training dataset (grouped by attributes, such as gender, race, religion, etc.) could be more vulnerable than others to the model inversion attacks. We also evaluate the risks incurred by model inversion attacks when the adversary does not have knowledge of all other non-sensitive attributes of the target record and demonstrate that our attack's performance is not impacted significantly in those scenarios. Finally, we empirically show that the MIAI attacks not only breach the privacy of a model's training data but also compromise distributional privacy.

Since the defense methods designed to mitigate reconstruction of instances resembling those used in the training dataset (TIR attacks)~\cite{alves2019,yang2020defending} do not directly apply to our MIAI attack setting, exploring new defense methods would be an interesting direction for future work. Moreover, defense mechanisms~\cite{FredriksonCCS2015} that perturb confidence scores but leave the model’s predicted labels unchanged are ineffective against our label-only attack. Therefore, designing effective defense methods that protect privacy against our label-only MIAI attack without degrading the target model's performance is left as future work.

%% file: Sections/Bib-Appendix.tex
\clearpage
\bibliographystyle{unsrt}
\bibliography{main}

\section*{Availability}
\label{sec:availability}

The original datasets, target models, attack model datasets, and attack models are (anonymously) available in the following links:
\begin{itemize}

\item GSS dataset: \\
\url{https://bigml.com/shared/dataset/gF5aUaBFNQ7QYNepUUg29a4Q2Lt}

Target models trained on GSS dataset:
\begin{itemize}
\item Decision tree model: \\
\url{https://bigml.com/shared/model/hBwXZNtvSBvJeRSLUxllA3wmrmU}
\item Deep neural network model: \\
\url{https://bigml.com/shared/deepnet/fx0ZgPycSuYr8QkUpezPCYMoRem}
\end{itemize}

\item Adult dataset: \\
\url{https://bigml.com/shared/dataset/l5DJvrXmPUnhBji9j8RrWpb7Mi6}

Target models trained on Adult dataset:
\begin{itemize}
\item Decision tree model: \\
\url{https://bigml.com/shared/model/1dI4W7rI8HB7yyWbUrWZzsAbZ95}
\item Deep neural network model: \\
\url{https://bigml.com/shared/deepnet/9HLcs6E9dveUHCL3Ca9pg92hPmx}
\end{itemize}

\item FiveThirtyEight dataset: \\
\url{https://bigml.com/shared/dataset/olFKJwZptAzdtugydSYza2TdDRN}

Target models trained on FiveThirtyEight dataset:
\begin{itemize}
\item Decision tree model: \\
\url{https://bigml.com/shared/model/oX9NQBIlzJ7q4p0TE9Z5zoPegNh}
\item Deep neural network model: \\
\url{https://bigml.com/shared/deepnet/3tk8ySX8J6VSqdWFYtBAAmuzLEr}
\end{itemize}
\end{itemize}

\begin{itemize}

\item Attack dataset obtained from the decision tree target model trained on GSS dataset and the corresponding ensemble attack model: \\
\begin{itemize}
\item Attack dataset: \\
\url{https://bigml.com/shared/dataset/wNguK1uWFsbFEXSMiODpdX4jlJc}
\item Ensemble attack model: \\
\url{https://bigml.com/shared/ensemble/9K9VffUC0ADjGmqROospSIAzY91}
\end{itemize}

\item Attack dataset obtained from the deep neural network target model trained on GSS dataset and the corresponding ensemble attack model: \\
\begin{itemize}
\item Attack dataset: \\
\url{https://bigml.com/shared/dataset/zu1hnA8nsECntgxMKa07mOVacnc}
\item Ensemble attack model: \\
\url{https://bigml.com/shared/ensemble/razFkSOUzaxeexpVDeGSlYSEXQu}
\end{itemize}

\item Attack dataset obtained from the decision tree target model trained on Adult dataset and the corresponding ensemble attack model: \\
\begin{itemize}
\item Attack dataset: \\
\url{https://bigml.com/shared/dataset/kvTpvptS1Hczj8Pgh4Iclr95h1m}
\item Ensemble attack model: \\
\url{https://bigml.com/shared/ensemble/jtAzcMkyIpFoXtfp6Rr8Ol6NNSi}
\end{itemize}

\item Attack dataset obtained from the deep neural network target model trained on Adult dataset  and the corresponding ensemble attack model: \\
\begin{itemize}
\item Attack dataset: \\
\url{https://bigml.com/shared/dataset/beAzpCmxYSwhvjIdqA9MvLJCgzo}
\item Ensemble attack model: \\
\url{https://bigml.com/shared/ensemble/danhxLiChOIC19qUfBBNXfv4FuM}
\end{itemize}

\item Attack dataset obtained from the decision tree target model trained on FiveThirtyEight dataset and the corresponding ensemble attack model: \\
\begin{itemize}
\item Attack dataset: \\
\url{https://bigml.com/shared/dataset/hjKe5C63b1cOoROW7ufs0QPWHPY}
\item Ensemble attack model: \\
\url{https://bigml.com/shared/ensemble/ikQ5bwBYPinGaI6ASAeu10RPvnM}
\end{itemize}

\item Attack dataset obtained from the deep neural network target model trained on FiveThirtyEight dataset  and the corresponding ensemble attack model: \\
\begin{itemize}
\item Attack dataset: \\
\url{https://bigml.com/shared/dataset/c2wVKvpEIlWfQveRqKncSzUEjlA}
\item Ensemble attack model: \\
\url{https://bigml.com/shared/ensemble/3QODgBv2xkSOc7qJzJ9ZYsEqTv6}
\end{itemize}
\end{itemize}

\appendix
\section{Appendix}

\begin{figure*}[h]
\centering
\subfigure[Marginal prior of the positive class attribute is $0.3$.]{\includegraphics[width=0.32\textwidth, height=5cm]
{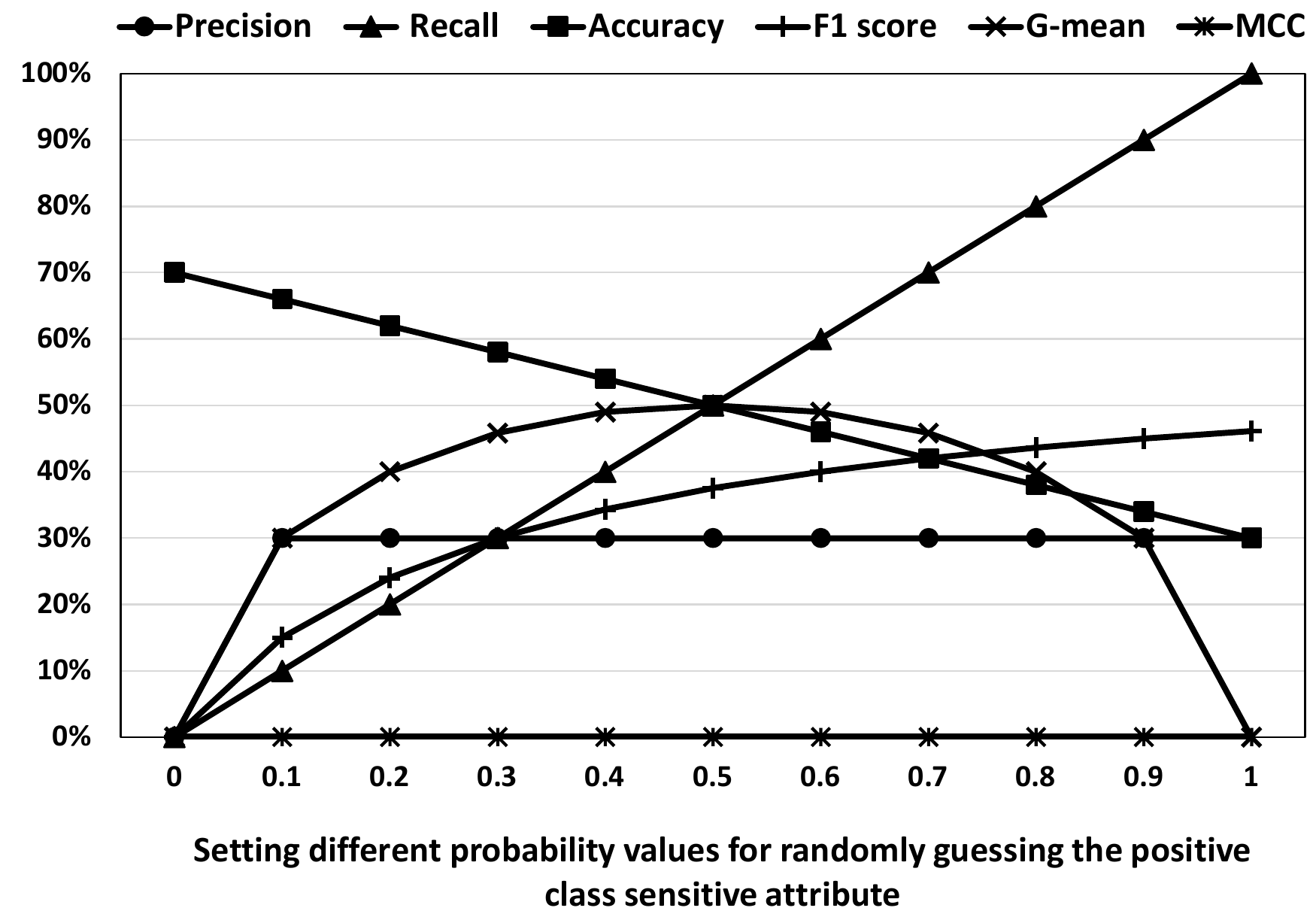}
\label{fig:random_30}
}
\hfill
\subfigure[GSS dataset where the marginal prior of the positive class attribute is $0.197$.]{\includegraphics[width=0.32\textwidth, height=5cm]
{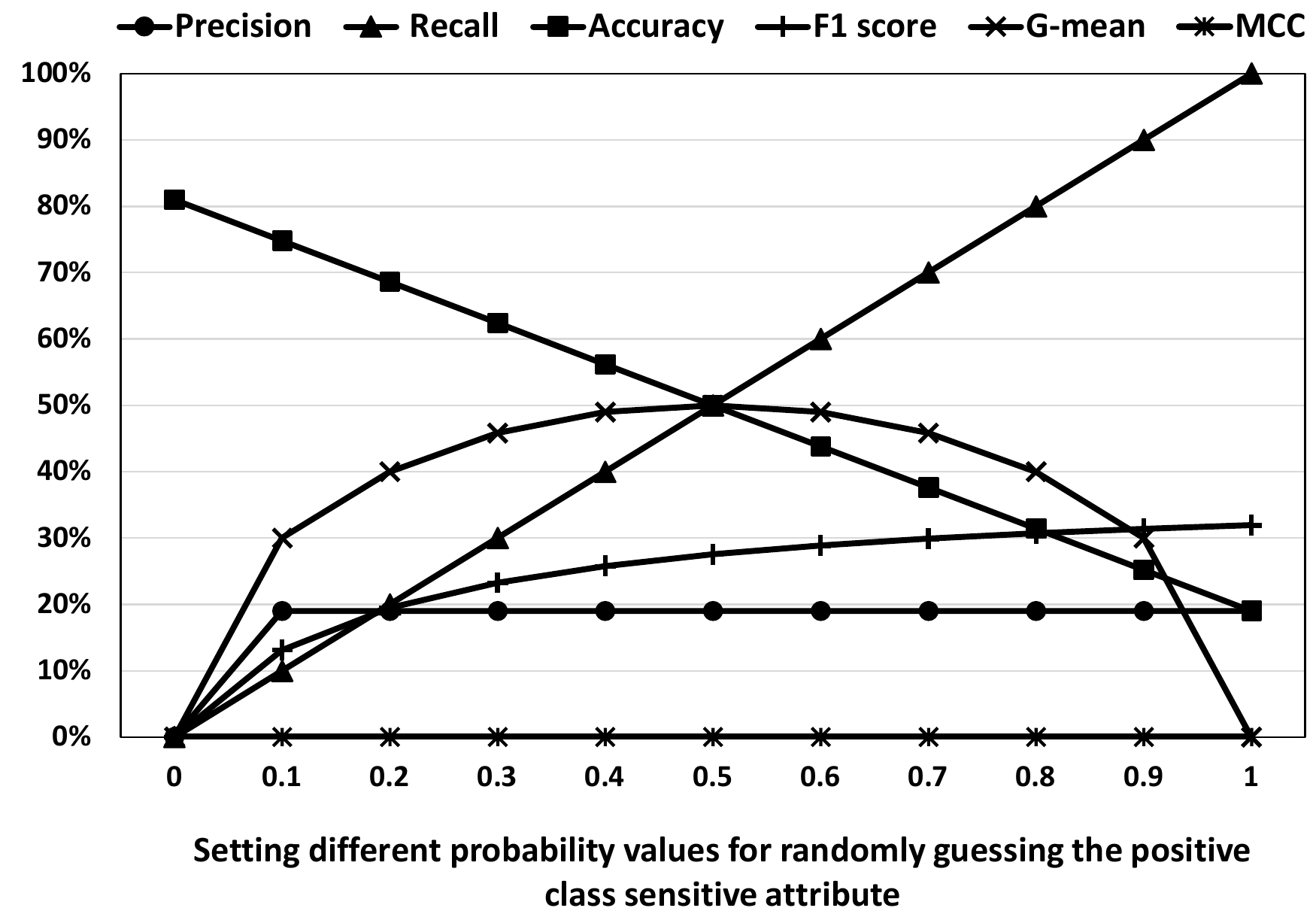}
\label{fig:random_gss}
}
\hfill
\subfigure[Adult dataset where the marginal prior of the positive class attribute is $0.479$.]{\includegraphics[width=0.32\textwidth, height=5cm]
{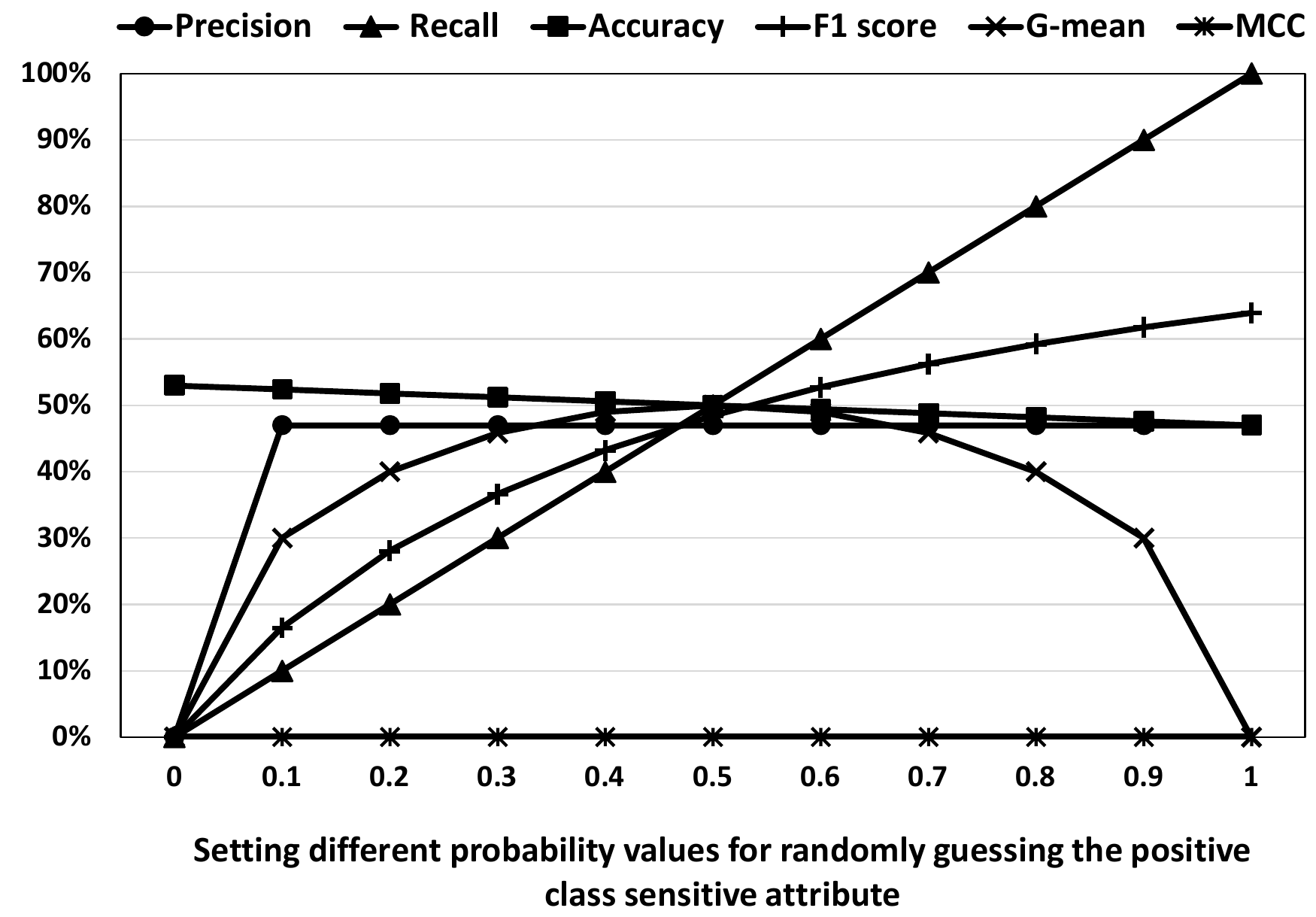}
\label{fig:random_adult}
}
\caption{Random guessing attack performances for different marginal priors of the positive class sensitive attribute value.}
\label{fig:random}
\end{figure*}

\begin{table*}[h]
  \centering
  \caption{Confusion matrix of decision tree target model trained on GSS dataset.}
  \vspace{-0.2cm}
  \resizebox{0.75\textwidth}{!}{
\begin{tabular}{|l|*{5}{c|}}\hline
    \backslashbox{Actual}{Predicted}
    &Not too happy& Pretty happy& Very happy & Total & Recall \\\hline
    Not too happy & 5 & 63 & 370 & 438 & 1.14\% \\\hline
    Pretty happy  & 0 & 813 & 4178 & 4991 & 16.29\% \\\hline
    Very happy    & 0 & 526 & 9280 & 9806 & 94.64\% \\\hline
    Total         & 5 & 1402 & 13828 & 15235 & Avg. recall 37.36\% \\ \hline
    Precision     & 100\% & 57.99\% & 67.11\% & Avg. precision 75.03\% & Accuracy 66.28\% \\ \hline
\end{tabular}
  }
\label{table:CM_DT_GSS}
\vspace{0.3cm}

  \centering
  \caption{Confusion matrix of deepnet target model trained on GSS dataset.}
  \resizebox{0.75\textwidth}{!}{
\begin{tabular}{|l|*{5}{c|}}\hline
    \backslashbox{Actual}{Predicted}
    &Not too happy& Pretty happy& Very happy & Total & Recall \\\hline
    Not too happy & 1 & 102 & 335 & 438 & 0.23\% \\\hline
    Pretty happy  & 0 & 565 & 4426 & 4991 & 11.32\% \\\hline
    Very happy    & 0 & 598 & 9208 & 9806 & 93.90\% \\\hline
    Total         & 1 & 1265 & 13969 & 15235 & Avg. recall 35.15\% \\ \hline
    Precision     & 100\% & 44.66\% & 65.92\% & Avg. precision 70.19\% & Accuracy 64.16\% \\ \hline
\end{tabular}
  }
\label{table:CM_DNN_GSS}
\vspace{0.3cm}

  \centering
  \caption{Confusion matrix of decision tree target model trained on Adult dataset.}
  \resizebox{0.6\textwidth}{!}{
\begin{tabular}{|l|*{4}{c|}}\hline
    \backslashbox{Actual}{Predicted}
    & <=50K & >50K  & Total & Recall \\\hline
    <=50K & 24912 & 1537 & 26449 & 94.19\% \\\hline
    >50K  & 3343 & 5430 & 8773 & 61.89\%  \\\hline
    Total & 28255 & 6967 & 35222  & Avg. recall 78.04\% \\ \hline
    Precision  & 88.17\% & 77.94\%  & Avg. precision 83.05\% & Accuracy 86.15\% \\ \hline
\end{tabular}
  }
\label{table:CM_DT_Adult}
\vspace{0.3cm}

  \centering
  \caption{Confusion matrix of deepnet target model trained on Adult dataset.}
  \resizebox{0.6\textwidth}{!}{
\begin{tabular}{|l|*{4}{c|}}\hline
    \backslashbox{Actual}{Predicted}
    & <=50K & >50K  & Total & Recall \\\hline
    <=50K & 24433 & 2016 & 26449 & 92.38\% \\\hline
    >50K  & 3276 & 5497 & 8773 & 62.66\%  \\\hline
    Total & 27709 & 7513 & 35222  & Avg. recall 77.52\% \\ \hline
    Precision  & 88.18\% & 73.17\%  & Avg. precision 80.67\% & Accuracy 84.97\% \\ \hline
\end{tabular}
  }
\label{table:CM_DNN_Adult}

\vspace{0.3cm}

  \centering
  \caption{Confusion matrix of decision tree target model trained on FiveThirtyEight dataset.}
  \resizebox{0.8\textwidth}{!}{
\begin{tabular}{|l|*{7}{c|}}\hline
    \backslashbox{Actual}{Predicted}
    &Medium & Medium Well & Medium Rare & Rare & Well & Total & Recall \\\hline
    Medium & 105 & 0 & 3 & 0 & 1 & 109 & 96.33\% \\\hline
    Medium Well  & 0 & 55 & 1 & 0 & 0 & 56 & 98.21\% \\\hline
    Medium Rare  & 3 & 1 & 122 & 1 & 1 & 128 & 95.31\% \\\hline
    Rare & 0 & 1 & 0 & 17 & 0 & 18 & 94.44\% \\\hline
    Well & 0 & 0 & 0 & 0 & 20 & 20 & 100\% \\\hline
    Total & 108 & 57 & 126 & 18 & 22 & 331 & Avg. Recall 96.86\% \\ \hline
    Precision & 97.22\% & 96.49\% & 96.83\% & 94.44\% & 90.91\% & Avg. Precision 95.18\% & Accuracy 96.37\% \\ \hline
\end{tabular}
}

\label{table:CM_DT_FTE}

\vspace{0.3cm}

  \centering
  \caption{Confusion matrix of deepnet target model trained on FiveThirtyEight dataset.}
  \resizebox{0.8\textwidth}{!}{
\begin{tabular}{|l|*{7}{c|}}\hline
    \backslashbox{Actual}{Predicted}
    &Medium & Medium Well & Medium Rare & Rare & Well & Total & Recall \\\hline
    Medium & 9 & 0 & 95 & 5 & 0 & 109 & 8.26\%  \\\hline
    Medium Well  & 10 & 0 & 42 & 4 & 0 & 56 & 0.00\%  \\\hline
    Medium Rare &  12 & 0 & 104 & 11 & 1 & 128 & 81.25\%  \\\hline
    Rare & 2 & 0 & 15 & 1 & 0 & 18 & 5.56\%  \\\hline
    Well & 3 & 0 & 13 & 4 & 0 & 20 & 0.00\%  \\\hline
    Total & 36 & 0 & 269 & 25 & 1 & 331 & Avg. Recall 19.01\%  \\ \hline
    Precision & 25.00\% & 0.00\% & 38.66\% & 4.00\% & 0.00\% & Avg. Precision 13.53\% & Accuracy 34.44\%  \\ \hline
\end{tabular}
}

\label{table:CM_DNN_FTE}

\end{table*}

\begin{table*}[t]
  \centering
  \caption{{Attack performance against the DT and DNN target models trained on GSS dataset.}}
  \resizebox{0.9\textwidth}{!}{
  \begin{tabular}{| l | l | l | l | l | l | l | l | l | l | l | l | l |}
    \hline
    Target Model & Attack Strategy &  TP & TN & FP & FN & Precision & Recall & Accuracy & F1 score & G-mean & MCC & FPR \\ \hline 
    DT/DNN & NaiveA & $0$ & $12218$ & $0$ & $3017$ & $0\%$ & $0\%$ & $80.2\%$ & $0\%$ & $0\%$ & $0\%$ & $0\%$\\ \hline 
    DT & FJRMIA~\cite{FredriksonCCS2015} & $131$ & $11709$ & $509$ & $2886$ & $20.47\%$ & $4.34\%$ & $77.72\%$ & $7.16\%$ & $20.39\%$ & $0.3\%$ & $4.17\%$\\ \hline 
    DT & CSMIA & $1490$ & $7844$ & $4373$ & $1528$ & $25.41\%$ & $49.37\%$ & $61.27\%$ & $33.55\%$ & $56.3\%$ & $11.1\%$ & $35.79\%$\\ \hline 
    DT & LOMIA & $1782$ & $5565$ & $6653$ & $1235$ & $21.13\%$ & $59.07\%$ & $48.22\%$ & $31.12\%$ & $51.87\%$ & $3.7\%$  & $54.45\%$\\ \hline 
    DNN & FJRMIA~\cite{FredriksonCCS2015} & $1$ & $12213$ & $5$ & $3016$ & $16.67\%$ & $0.03\%$ & $80.17\%$ & $0.07\%$ & $1.82\%$ & $-0.2\%$ & $0.04\%$\\ \hline 
    DNN & CSMIA & $1212$ & $8058$ & $4160$ & $1805$ & $22.56\%$ & $40.17\%$ & $60.85\%$ & $28.89\%$ & $51.47\%$ & $5.1\%$ & $34.05\%$\\ \hline 
    DNN& LOMIA & $1225$ & $8015$ & $4203$ & $1792$ & $22.57\%$ & $40.6\%$ & $60.65\%$ & $29.01\%$ & $51.61\%$ & $5.16\%$  & $34.4\%$\\ \hline

  \end{tabular}
  }

\label{table:gss_results_dt_dnn}

\end{table*}

\begin{table*}[h]
  \centering
  \caption{Our proposed attacks' performance details against the decision tree target model trained on GSS dataset.}
  \resizebox{.95\textwidth}{!}{
  \begin{tabular}{ | l | l | l | l | l | l |l | l | l | l | l | l |}
    \hline
    Attack & Case &  TP & TN & FP & FN & Precision & Recall & Accuracy & F1 score & G-mean & MCC \\ \hline 
    Confidence score-based attack &  \multirow{2}{*}{(1)} & $219$ & $1336$ & $698$ & $134$ & $23.88\%$ & $62.04\%$ & $65.14\%$ & $34.49\%$ & $63.83\%$ & $20.2\%$ \\ 
    \cline{1, 3-12}    Label-only attack &   & $219$ & $1337$ & $697$ & $134$ & $23.91\%$ & $62.04\%$ & $65.19\%$ & $34.52\%$ & $63.86\%$ & $20.3\%$ \\ \hline
    Confidence score-based attack & \multirow{2}{*}{(2)}  & $661$ & $4466$ & $2409$ & $1007$ & $21.53\%$ & $39.63\%$ & $60.01\%$ & $27.91\%$ & $50.74\%$ & $3.8\%$ \\
    \cline{1, 3-12}    Label-only attack &              & $1227$ & $1848$ & $5028$ & $440$ & $19.61\%$ & $73.61\%$ & $36\%$ & $30.98\%$ & $44.48\%$ & $0.4\%$ \\ \hline
    Confidence score-based attack & \multirow{2}{*}{(3)} & $610$ & $2042$ & $1266$ & $387$ & $32.52\%$ & $61.18\%$ & $61.61\%$ & $42.46\%$ & $61.46\%$ & $19.5\%$ \\
    \cline{1, 3-12}    Label-only attack &             & $336$ & $2380$ & $928$ & $661$ & $26.58\%$ & $33.7\%$ & $63.09\%$ & $29.72\%$ & $49.24\%$ & $5.2\%$ \\ \hline 
  \end{tabular}
  }
\label{table:gss_dt_bpmpcs_details}

  \centering
  \caption{Our proposed attacks' performance details against the deep neural network target model trained on GSS dataset.}
  \vspace{0.2cm}
  \resizebox{.95\textwidth}{!}{
  \begin{tabular}{ | l | l | l | l | l | l |l | l | l | l | l | l |}
    \hline
    Attack & Case &  TP & TN & FP & FN & Precision & Recall & Accuracy & F1 score & G-mean & MCC \\ \hline 
    Confidence score-based attack & \multirow{2}{*}{(1)}  & $96$ & $468$ & $317$ & $130$ & $23.24\%$ & $42.48\%$ & $55.79\%$ & $30.05\%$ & $50.32\%$ & $1.8\%$ \\ 
    \cline{1, 3-12} Label-only attack & & $96$ & $469$ & $316$ & $130$ & $23.3\%$ & $42.48\%$ & $55.89\%$ & $30.09\%$ & $50.38\%$ & $1.9\%$ \\ \hline
    Confidence score-based attack & \multirow{2}{*}{(2)}  & $55$ & $7339$ & $205$ & $1611$ & $21.15\%$ & $3.3\%$ & $80.28\%$ & $5.71\%$ & $17.92\%$ & $1.4\%$ \\ 
    \cline{1, 3-12}  Label-only attack &  & $94$ & $7166$ & $378$ & $1572$ & $19.92\%$ & $5.64\%$ & $78.83\%$ & $8.79\%$ & $23.15\%$ & $1.1\%$ \\ \hline
    Confidence score-based attack & \multirow{2}{*}{(3)} & $1061$ & $251$ & $3638$ & $64$ & $22.58\%$ & $94.31\%$ & $26.17\%$ & $36.44\%$ & $24.67\%$ & $1.3\%$ \\ 
    \cline{1, 3-12}  Label-only attack &  & $1035$ & $380$ & $3509$ & $90$ & $22.78\%$ & $92\%$ & $28.22\%$ & $36.51\%$ & $29.98\%$ & $2.5\%$ \\ \hline 

  \end{tabular}
  }
\label{table:gss_dnn_bpmpcs_details}
\end{table*}

\begin{table*}[h]
  \centering
  \caption{{Attack performance against the DT and DNN target models trained on  Adult dataset.}}
  \resizebox{0.9\textwidth}{!}{
  \begin{tabular}{ | l | l | l | l | l | l | l | l | l | l | l | l | l |}
    \hline
    Target Model & Attack Strategy &  TP & TN & FP & FN & Precision & Recall & Accuracy & F1 score & G-mean & MCC & FPR \\ \hline
    DT/DNN & NaiveA & $0$ & $18329$ & $0$ & $16893$ & $0\%$ & $0\%$ & $52.04\%$ & $0\%$ & $0\%$ & $0\%$ & $0\%$\\ \hline 
    DT & FJRMIA~\cite{FredriksonCCS2015} & $3788$ & $17818$ & $511$ & $13105$ & $88.11\%$ & $22.42\%$ & $61.34\%$ & $35.75\%$ & $46.69\%$ & $29.9\%$ & $2.79\%$\\ \hline 
    DT & CSMIA & $7664$ & $17085$ & $1244$ & $9229$ & $86.04\%$ & $45.37\%$ & $70.27\%$ & $59.41\%$ & $65.03\%$ & $44.3\%$ & $6.79\%$\\ \hline 
    DT & LOMIA & $7574$ & $17132$ & $1197$ & $9319$ & $86.35\%$ & $44.84\%$ & $70.14\%$ & $59.02\%$ & $64.74\%$ & $44.3\%$  & $6.53\%$\\ \hline
    DNN & FJRMIA~\cite{FredriksonCCS2015} & $3592$ & $17717$ & $612$ & $13301$ & $85.44\%$ & $21.26\%$ & $60.5\%$ & $34.05\%$ & $45.34\%$ & $27.6\%$ & $3.34\%$\\ \hline 
    DNN & CSMIA & $7490$ & $17139$ & $1190$ & $9403$ & $86.29\%$ & $44.34\%$ & $69.93\%$ & $58.58\%$ & $64.39\%$ & $43.9\%$ & $6.49\%$\\ \hline 
    DNN & LOMIA & $7565$ & $17121$ & $1208$ & $9328$ & $86.23\%$ & $44.78\%$ & $70.09\%$ & $58.95\%$ & $64.68\%$ & $44.2\%$  & $6.59\%$\\ \hline
    
  \end{tabular}
  }
\label{table:adult_results_dt_dnn}

\end{table*}

\begin{table*}[h]

  \centering
  \caption{Our proposed attacks' performance details against the decision tree target model trained on Adult dataset.}
  \resizebox{.95\textwidth}{!}{
  \begin{tabular}{ | l | l | l | l | l | l |l | l | l | l | l | l |}
    \hline
    Attack & Case &  TP & TN & FP & FN & Precision & Recall & Accuracy & F1 score & G-mean & MCC \\ \hline
    Confidence score-based attack & \multirow{2}{*}{(1)}  & $3788$ & $3466$ & $511$ & $1498$ & $88.11\%$ & $71.66\%$ & $78.31\%$ & $79.04\%$ & $79.03\%$ & $58.4\%$ \\
    \cline{1, 3-12} Label-only attack &   & $3787$ & $3466$ & $511$ & $1499$ & $88.11\%$ & $71.64\%$ & $78.3\%$ & $79.03\%$ & $79.02\%$ & $58.4\%$ \\ \hline
    Confidence score-based attack & \multirow{2}{*}{(2)}  & $1375$ & $13560$ & $456$ & $7697$ & $75.09\%$ & $15.16\%$ & $64.68\%$ & $25.22\%$ & $38.29\%$ & $21.5\%$ \\
    \cline{1, 3-12} Label-only attack &   & $1275$ & $13626$ & $390$ & $7797$ & $76.58\%$ & $14.05\%$ & $64.54\%$ & $23.75\%$ & $36.96\%$ & $21.3\%$ \\ \hline
    Confidence score-based attack & \multirow{2}{*}{(3)} & $2501$ & $59$ & $277$ & $34$ & $90.03\%$ & $98.66\%$ & $89.17\%$ & $94.15\%$ & $41.62\%$ & $29.5\%$ \\
    \cline{1, 3-12} Label-only attack & & $2512$ & $40$ & $296$ & $23$ & $89.46\%$ & $99.09\%$ & $88.89\%$ & $94.03\%$ & $34.35\%$ & $24.1\%$ \\ \hline

  \end{tabular}
  }
\label{table:adult_dt_bpmpcs_details}

  \centering
  \caption{Our proposed attacks' performance details against the deep neural network target model trained on Adult dataset.}
\vspace{0.2cm}
  \resizebox{.95\textwidth}{!}{
  \begin{tabular}{ | l | l | l | l | l | l |l | l | l | l | l | l |}
    \hline
    Attack & Case &  TP & TN & FP & FN & Precision & Recall & Accuracy & F1 score & G-mean & MCC \\ \hline 
    Confidence score-based attack &  \multirow{2}{*}{(1)} & $3592$ & $3838$ & $612$ & $1918$ & $85.44\%$ & $65.19\%$ & $74.6\%$ & $73.96\%$ & $74.98\%$ & $51.8\%$ \\
    \cline{1, 3-12} Label-only attack &   & $3592$ & $3838$ & $612$ & $1918$ & $85.44\%$ & $65.19\%$ & $74.6\%$ & $73.96\%$ & $74.98\%$ & $51.8\%$ \\ \hline
    Confidence score-based attack & \multirow{2}{*}{(2)}  & $1467$ & $13235$ & $344$ & $7454$ & $81.01\%$ & $16.44\%$ & $65.34\%$ & $27.34\%$ & $40.03\%$ & $25\%$ \\
    \cline{1, 3-12} Label-only attack &   & $1542$ & $13216$ & $363$ & $7379$ & $80.94\%$ & $17.29\%$ & $65.59\%$ & $28.49\%$ & $41.02\%$ & $25.7\%$ \\ \hline
    Confidence score-based attack & \multirow{2}{*}{(3)} & $2431$ & $66$ & $234$ & $31$ & $91.22\%$ & $98.74\%$ & $90.41\%$ & $94.83\%$ & $46.61\%$ & $35.1\%$ \\
    \cline{1, 3-12} Label-only attack & & $2431$ & $67$ & $233$ & $31$ & $91.25\%$ & $98.74\%$ & $90.44\%$ & $94.85\%$ & $46.96\%$ & $35.4\%$ \\ \hline
  \end{tabular}
  }
\label{table:adult_dnn_bpmpcs_details}
\end{table*}

\subsection{Random Guessing Attack Performances}
\label{appn:random}
In this attack, the adversary randomly predicts the sensitive attribute by setting a probability for the positive class sensitive attribute value. Fig.~\ref{fig:random_30} shows the optimal performance of random guessing attack when the marginal prior of the positive class sensitive attribute is $0.3$ and the adversary sets different probabilities to predict the positive class sensitive attribute value (probabilities in x-axis). As shown in the figure, the maximum G-mean a random guessing attack can achieve is 50\%, independent of the knowledge of marginal prior. The precision for predicting the positive class sensitive attribute is constant and equals the marginal prior of that class as long as the  set probability is $>0$. This is because when the attack randomly assigns positive class label to the records, approximately 30\% of those records' sensitive attributes would turn out to be originally positive according to the marginal prior of the positive class sensitive attribute which is $0.3$. The recall of random guessing attack increases with the probability set to predict the positive class sensitive attribute. For example, if the adversary reports all the records' sensitive attributes as positive, there is no false negative left and thus recall reaches 100\%. Figures~\ref{fig:random_gss} and~\ref{fig:random_adult} show the performance of random guessing attack on the GSS and Adult datasets, respectively, when the adversary sets different probability values to predict the positive class sensitive attribute.
As shown in Figures~\ref{fig:random_30},~\ref{fig:random_gss}, and~\ref{fig:random_adult}, the MCC of the random guessing attacks is always 0.

\subsection{CSMIA  With Partial Knowledge of Non-sensitive Attributes}
\label{appn:CSMIA_partial}

For simplicity, we assume that there is only one non-sensitive attribute that is unknown to the adversary. Extending our attack steps to more than one unknown attribute is straightforward.
Without loss of generality, let $x_2 \in \mathbf{x}$ be the non-sensitive attribute unknown to the adversary.

Let $u$ be the number of unique possible values of $x_2$. We query the model by varying the unknown non-sensitive attribute with its different unique possible values (in the same way we vary the sensitive attribute $x_1$ in the attacks described in Section~\ref{sec:new_attacks}) while all other known non-sensitive attributes $\{x_3, ..., x_d\}$ remain the same. {When the non-sensitive attributes are continuous, we use binning to put them into categories just like we did for sensitive attributes}. Hence, in this attack, we query the model $u$ times for each possible value of the sensitive attribute. As a result, the complexity of the attacks described in this section is $u$ times the complexity of the attacks in Section~\ref{sec:new_attacks}. 

According to the notations used in Section~\ref{sec:new_attacks}, let $C_{0}$=${\textstyle\sum}_{i=1}^{u} (y = y'_{0\_i})$ be the number of times the predictions are correct with the sensitive attribute \emph{no} and $C_{1}$=${\textstyle\sum}_{i=1}^{u} (y = y'_{1\_i})$  be the number of times the predictions are correct with the sensitive attribute \emph{yes}.

In order to determine the value of $x_1$, this attack considers the following cases:
    \begin{enumerate}[label=\fbox{\textbf{Case (\arabic*)}},   leftmargin=0pt, itemindent=53pt]
        \item If $C_{0}$ != $C_{1}$, i.e., the number of correct target model predictions are different for different sensitive attribute values, the attack selects the sensitive attribute to be the one for which the number of correct predictions is higher. For instance, if $C_{1}>$  $C_{0}$, the attack predicts $yes$ for the sensitive attribute and vice versa. 
        \vspace{0.1cm}
        \item If $C_{0}$ = $C_{1}$ and both are non-zero, we compute the sum of the confidence scores (only for the correct predictions) for each sensitive attribute and the attack selects the sensitive attribute to be the one for which the sum of the confidence scores is the maximum. 
        \vspace{0.1cm}
        \item If $C_{0} = 0$ $\wedge$ $C_{1} = 0 $, we compute the sum of the confidence scores for each sensitive attribute and the attack selects the sensitive attribute to be the one for which the sum of the confidence scores is the minimum.
    \end{enumerate}
If there is a second non-sensitive attribute that is unknown to the adversary (let that unknown attribute be $x_3$) and $v$ is the number of unique possible values for that unknown non-sensitive attribute, we query the model by varying both $x_2$ and $x_3$ while all other known non-sensitive attributes $\{x_4, ..., x_d\}$ remain the same. Hence, in this attack, we query the model $u * v$ times for each possible value of the sensitive attribute. As a result, the complexity of the attack becomes $u * v$ times the complexity of the attacks in Section~\ref{sec:new_attacks}.

\begin{figure}[t]
\centering
\includegraphics[width=0.47\textwidth, height=4.1cm]
{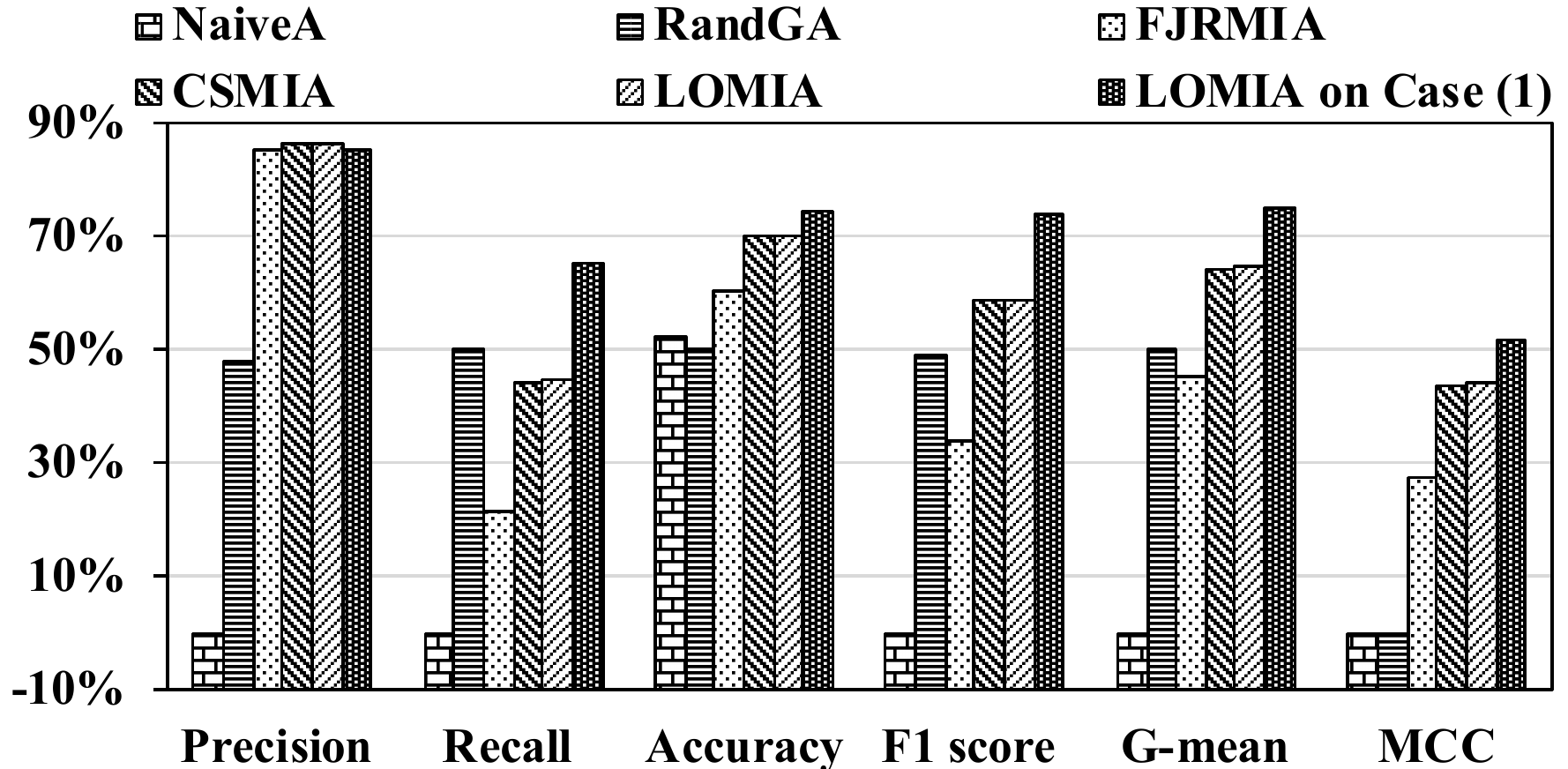}
\caption{Comparison of attacks: FJRMIA~\cite{FredriksonCCS2015}, CSMIA, and LOMIA with baseline attack strategies NaiveA and RandGA against Deepnet model trained on Adult dataset.}
\label{fig:Adult_RandomDNN_NEW}
\end{figure}

\subsection{Disparate Vulnerability of Model Inversion Attack}
\label{appnC}
\begin{figure*}[h]
\centering
\includegraphics[width=0.97\textwidth, height=3.5cm]
{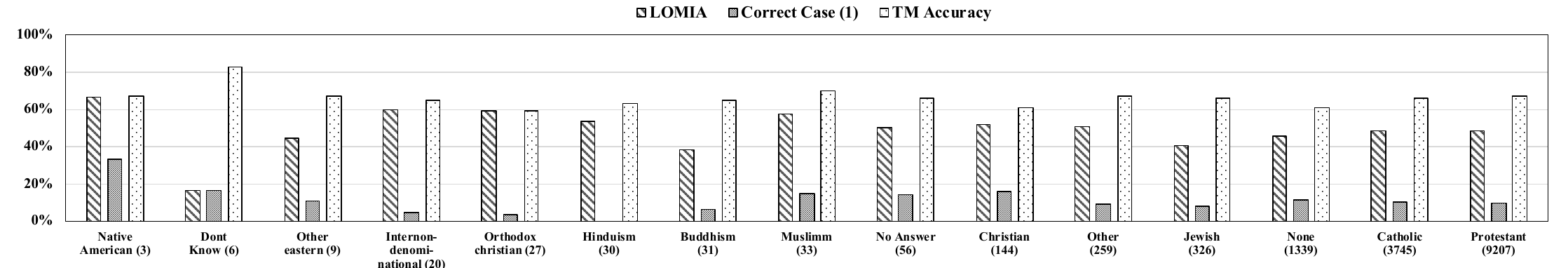}
\caption{Disparate vulnerability of LOMIA for different religion groups. The results represent attack on the decision tree target model trained on GSS dataset.}
\label{fig:disp_gss_rel}
\end{figure*}

\begin{figure*}[h]
\centering
\includegraphics[width=0.97\textwidth, height=3.5cm]
{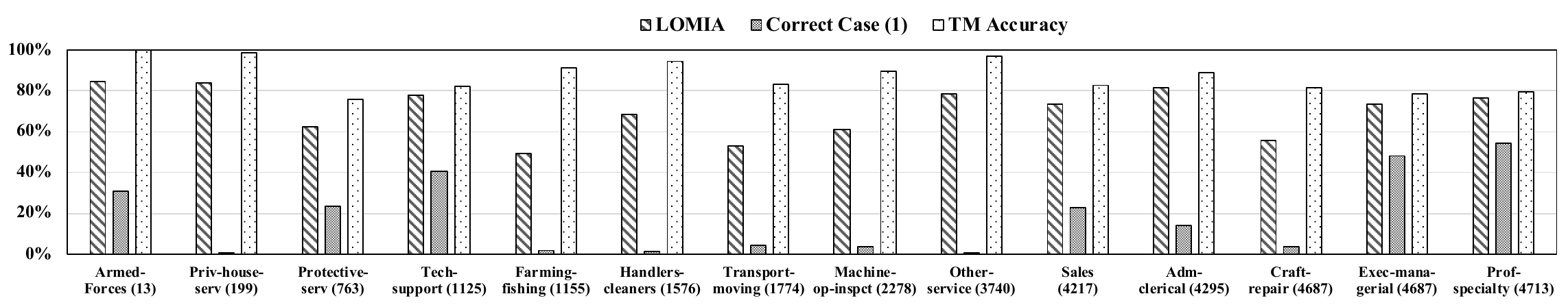}
\caption{Disparate vulnerability of LOMIA for different occupation groups. The results represent attack on the deepnet target model trained on Adult dataset.}
\label{fig:disp_adult_occ}
\end{figure*}

Fig.~\ref{fig:disp_gss_rel} and~\ref{fig:disp_adult_occ} show the contrast in the  performances of LOMIA against different religion and occupation populations. The attacks are performed against the decision tree model trained on GSS dataset and the deepnet model trained on Adult dataset, respectively. The x-axis represents religion and occupation populations along with the number of records in the training dataset that belong to the particular subgroups. The results show that certain religion and occupation subgroups are more vulnerable to model inversion attacks than others.

\subsection{CSMIA Results With Partial Knowledge of Target Record's Non-sensitive Attributes}
\label{appnB:CSMIA_partial}
\begin{figure*}[t]
\centering
\includegraphics[width=1\textwidth, height=3.8cm]
{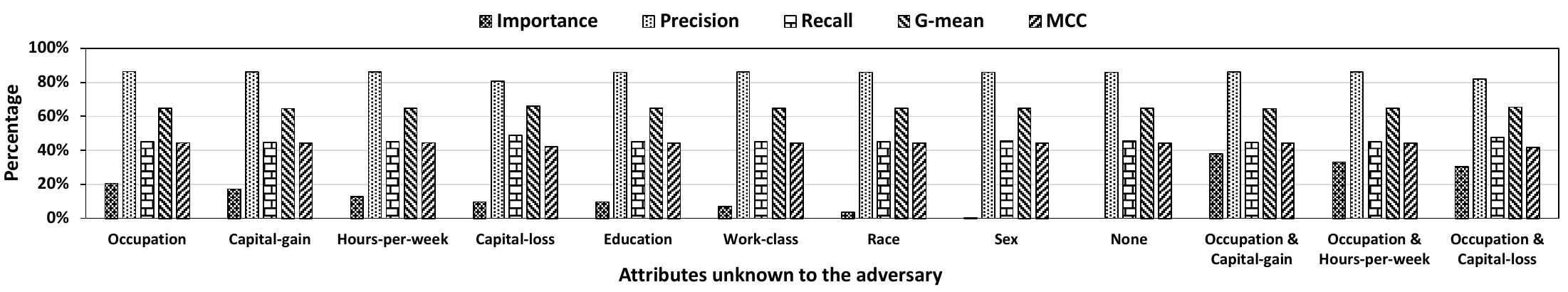}
\caption{CSMIA performance against the decision tree model trained on Adult dataset when some of the other (non-sensitive) attributes of a target individual are also unknown to the adversary.}
\label{fig:missing_adult_dt_cs}
\end{figure*}

Excluding the sensitive attribute (`marital status') and the output of the target model (`income'), we first consider each of the remaining (non-sensitive) attributes to be unknown to the adversary once at a time, i.e., denoting those as $x_2$. Figure~\ref{fig:missing_adult_dt_cs} shows the performance of CSMIA on the decision tree target model trained on the Adult dataset when some of the non-sensitive attributes are unknown to the adversary. The x-axis shows the non-sensitive attributes that are unknown. The attributes are sorted (from left to right) according to their \emph{importance} in the model, a parameter computed by BigML. We also present the original results (i.e., when \emph{none} of the non-sensitive attributes is unknown to the adversary) to compare how the partial knowledge of the target individual's non-sensitive attributes impacts our attacks' performances. As demonstrated in Figure~\ref{fig:missing_adult_dt_cs}, we observe that the performance of our attack does not deteriorate and remains almost the same when some of the non-sensitive attributes are unknown to the adversary, independent of the importance of the attributes in the target model. We observe only slightly lower precision (and slightly higher recall) when the `capital-loss' attribute is unknown to the adversary. 
We also perform experiments where a combination of non-sensitive attributes are unknown to the adversary-- `occupation and capital-gain' (combined importance $37.8\%$),  `occupation and hours-per-week' (combined importance $33.3\%$), and `occupation and capital-loss' (combined importance $30.4\%$). As demonstrated in Figure~\ref{fig:missing_adult_dt_cs}, our attack does not show any significant deterioration. 
{Table~\ref{table:CSMIA_queries} shows the number of queries to the target model for the above experiments.
Due to the combinatorial complexity of our CSMIA partial knowledge attack, we limit the number of unknown non-sensitive attributes to two for these experiments. 
}

\begin{figure*}[t]
\centering
\includegraphics[width=0.9\textwidth, height=3.8cm]
{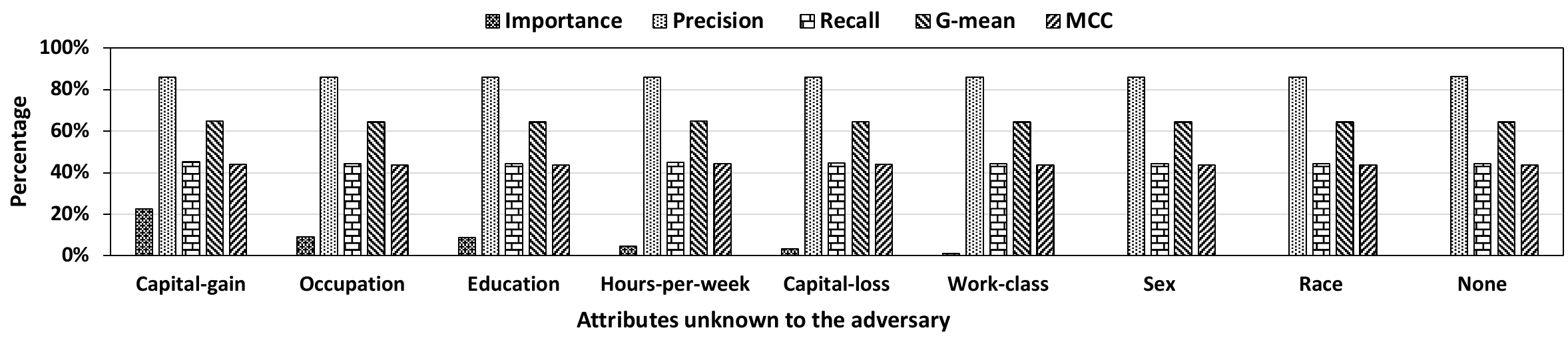}
\caption{CSMIA performance against the deep neural network model trained on Adult dataset when some of the other (non-sensitive) attributes of a target individual are also unknown to the adversary.}
\label{fig:missing_adult_dnn_cs}
\end{figure*}

Fig.~\ref{fig:missing_adult_dnn_cs} shows the performance of our confidence score-based attack on the deep neural network target model trained on the Adult dataset when some of the non-sensitive attributes are unknown to the adversary. The x-axis shows the non-sensitive attribute that is unknown. The attributes are sorted (from left to right) according to their \emph{importance} in the model. We also present the original results (i.e., when \emph{none} of the non-sensitive attributes is unknown to the adversary) to compare how the partial knowledge of the target individual's non-sensitive attributes impacts our attacks' performances. 
As demonstrated in the figure, we observe that the performances of our attack do not deteriorate and remain almost the same when some of the non-sensitive attributes are unknown to the adversary, independent of the importance of the attributes in the target model.

\begin{figure*}[h]
\centering
\subfigure[GSS dataset attributes' importance]{\includegraphics[width=0.45\textwidth, height=4cm]
{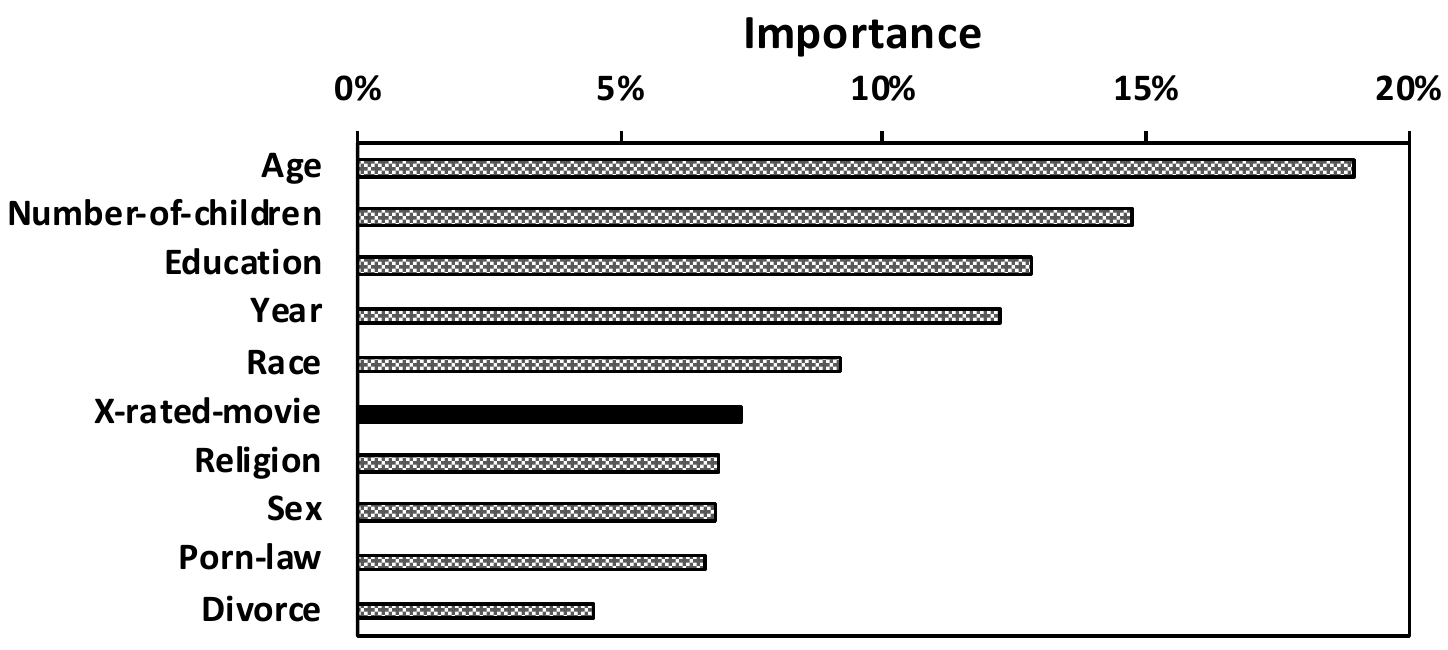}
}
\hfill
\subfigure[Adult dataset attributes' importance]{\includegraphics[width=0.45\textwidth, height=4cm]
{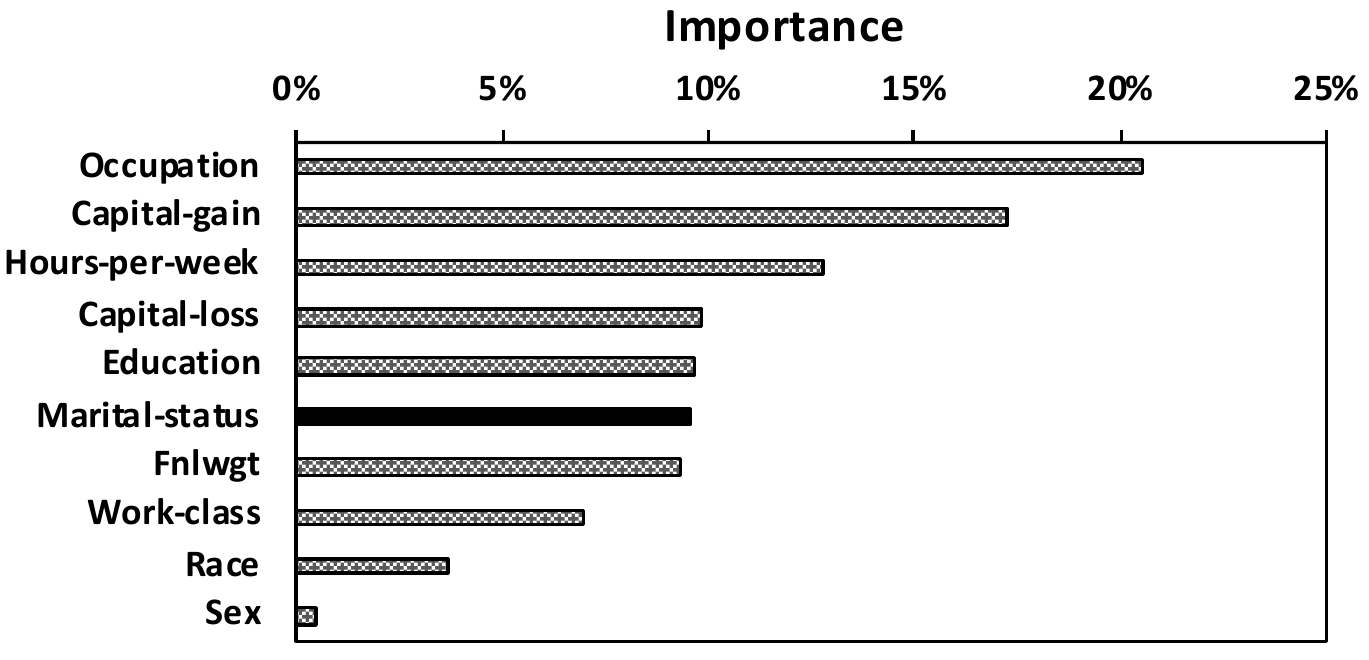}
}
\caption{Importance of GSS and Adult dataset attributes in their corresponding decision tree target models.}
\label{fig:importance}
\end{figure*}

\begin{figure*}[t]
\centering
\subfigure[]
{\includegraphics[width=0.3\textwidth, height=3.5cm]
{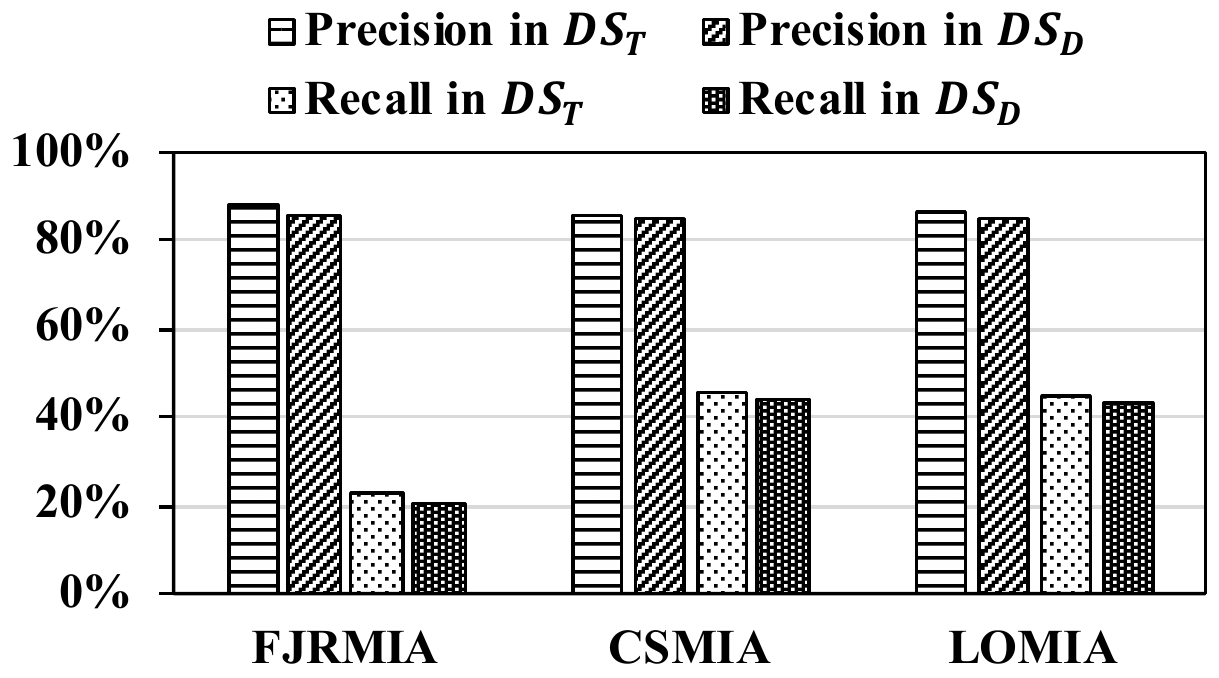}
\label{fig:dis_privacy}
}
\hfill
\subfigure[]
{\includegraphics[width=0.3\textwidth, height=3.5cm]
{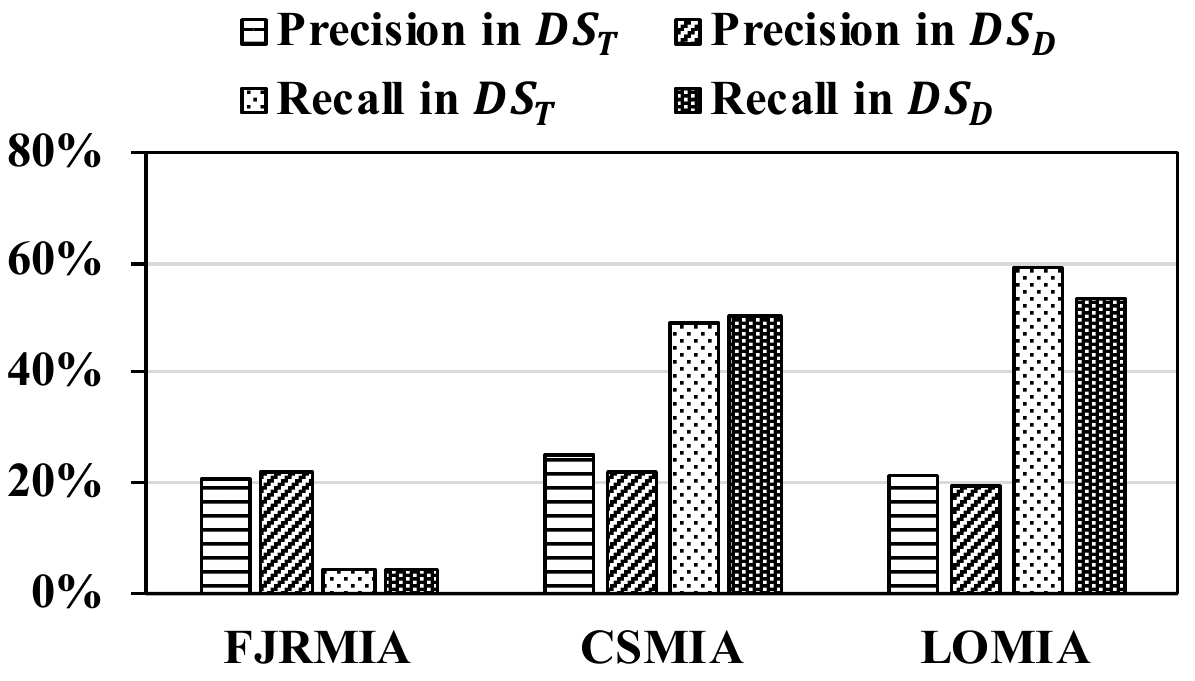}
\label{fig:dis_privacy}
}
\hfill
\subfigure[]
{\includegraphics[width=0.3\textwidth, height=3.5cm]
{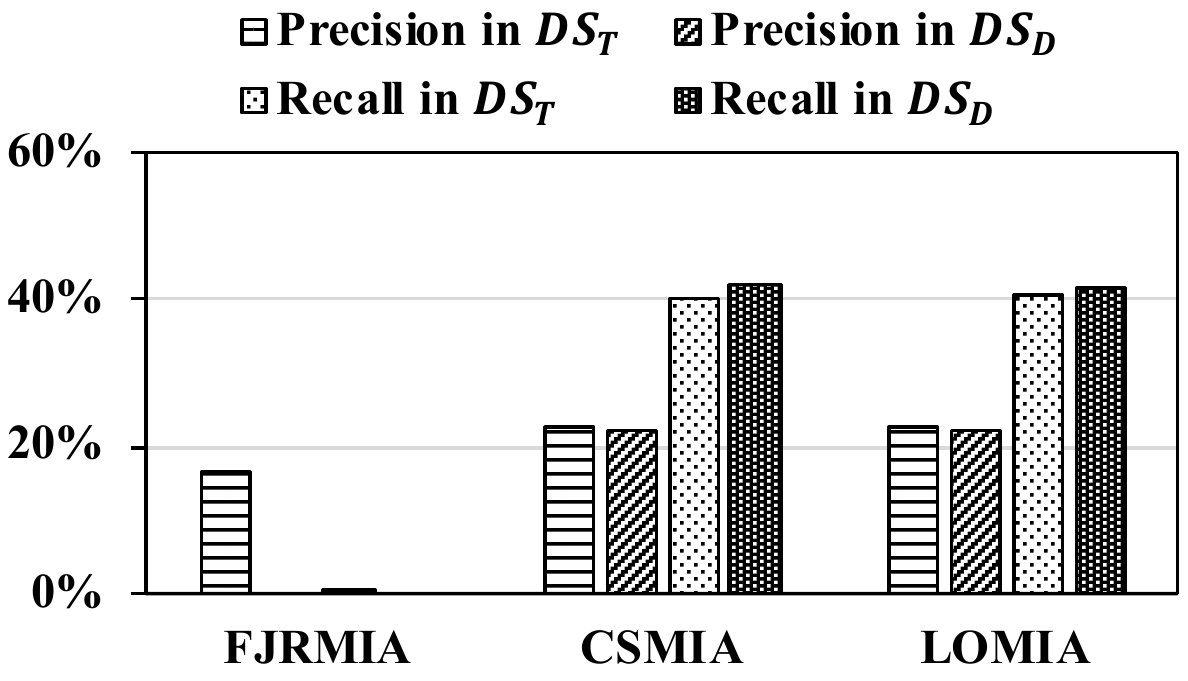}
\label{fig:dis_privacy}
}
\hfill
\caption{{Privacy leakage for $DS_T$ and $DS_D$: against (a) deepnet target model trained on Adult dataset, (b) decision tree target model trained on GSS dataset, and (c) deepnet target model trained on GSS dataset.}}
\label{fig:distributional_p}
\end{figure*}

\begin{figure*}[t]
\centering
\subfigure[Adult dataset attributes' importance in the LOMIA attack model trained against the decision tree target model]{\includegraphics[width=0.45\textwidth, height=4cm]
{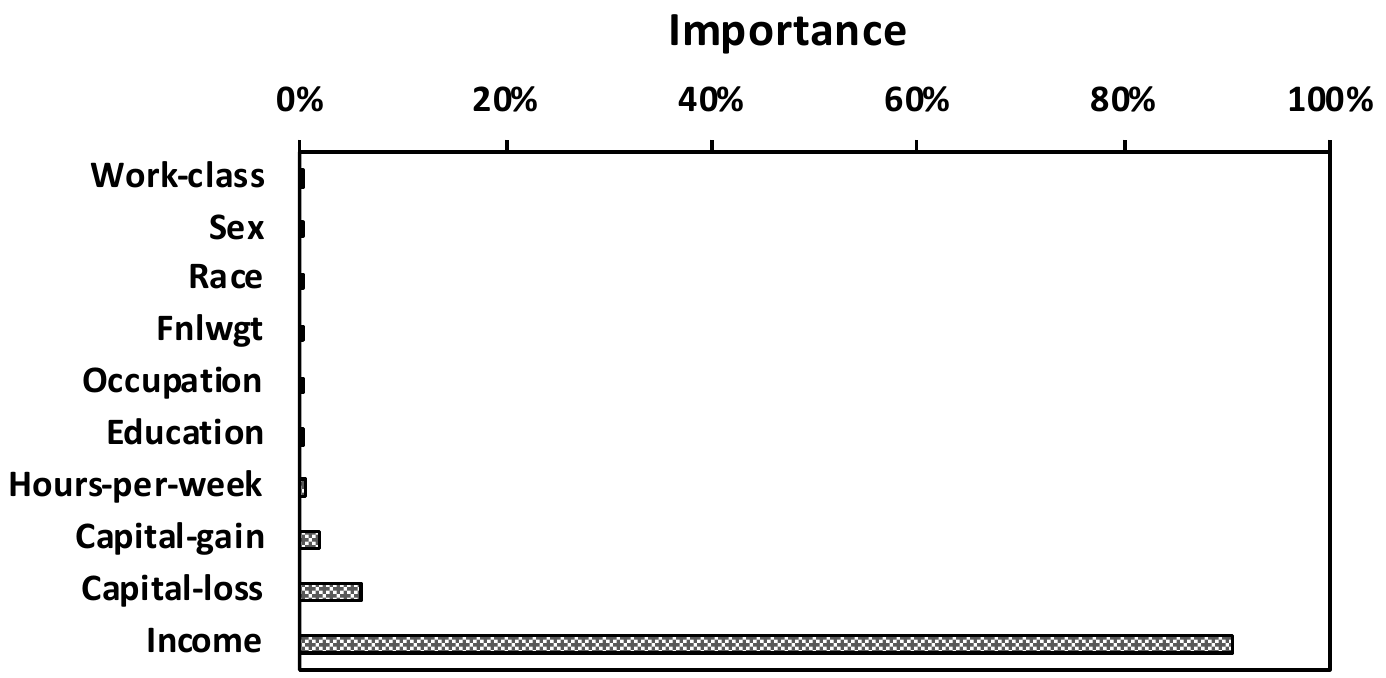}
}
\hfill
\subfigure[Adult dataset attributes' importance in the LOMIA attack model trained against the deepnet target model]{\includegraphics[width=0.45\textwidth, height=4cm]
{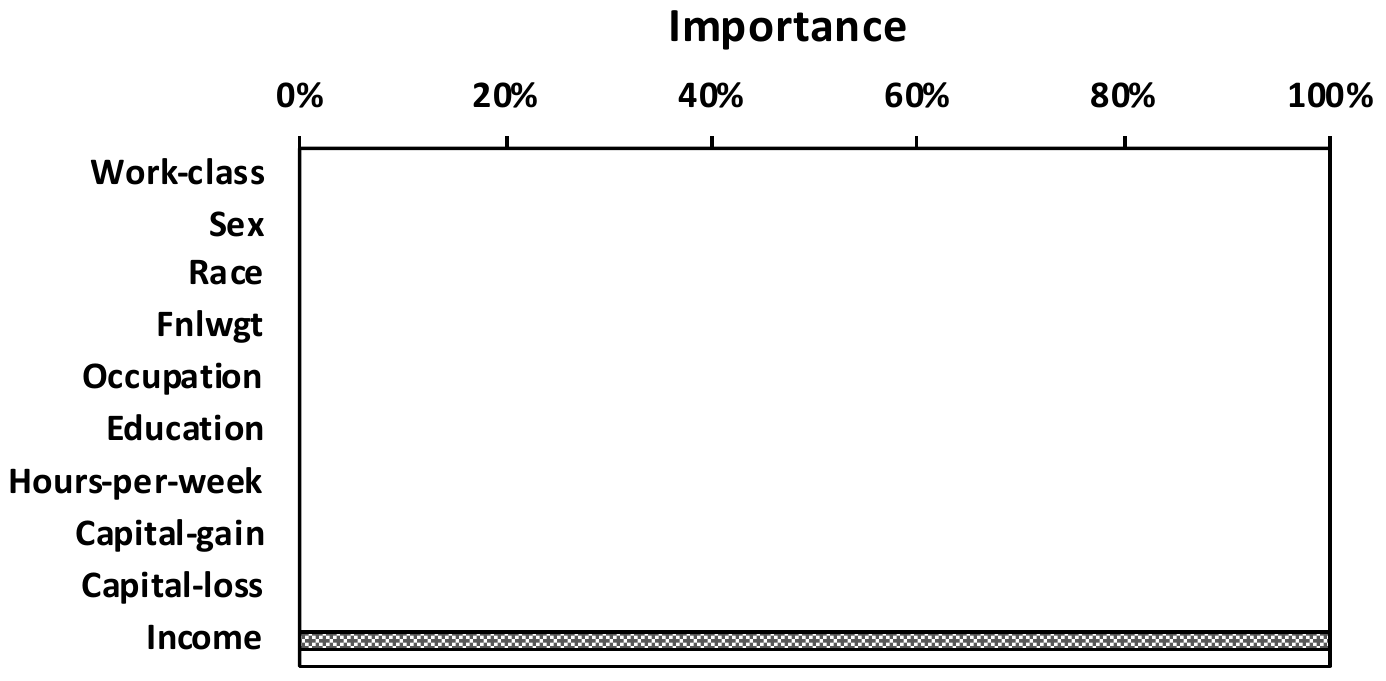}
}
\caption{{Importance of Adult dataset attributes in the LOMIA attack models trained against the decision tree and deepnet target models, respectively. Note that, the income attribute occupies 100\% importance in the LOMIA attack model trained against the deepnet target model.}}
\label{fig:importance_adult}
\end{figure*}

\begin{figure*}[t]
\centering
\subfigure[GSS dataset attributes' importance in the LOMIA attack model trained against the decision tree target model]{\includegraphics[width=0.45\textwidth, height=4cm]
{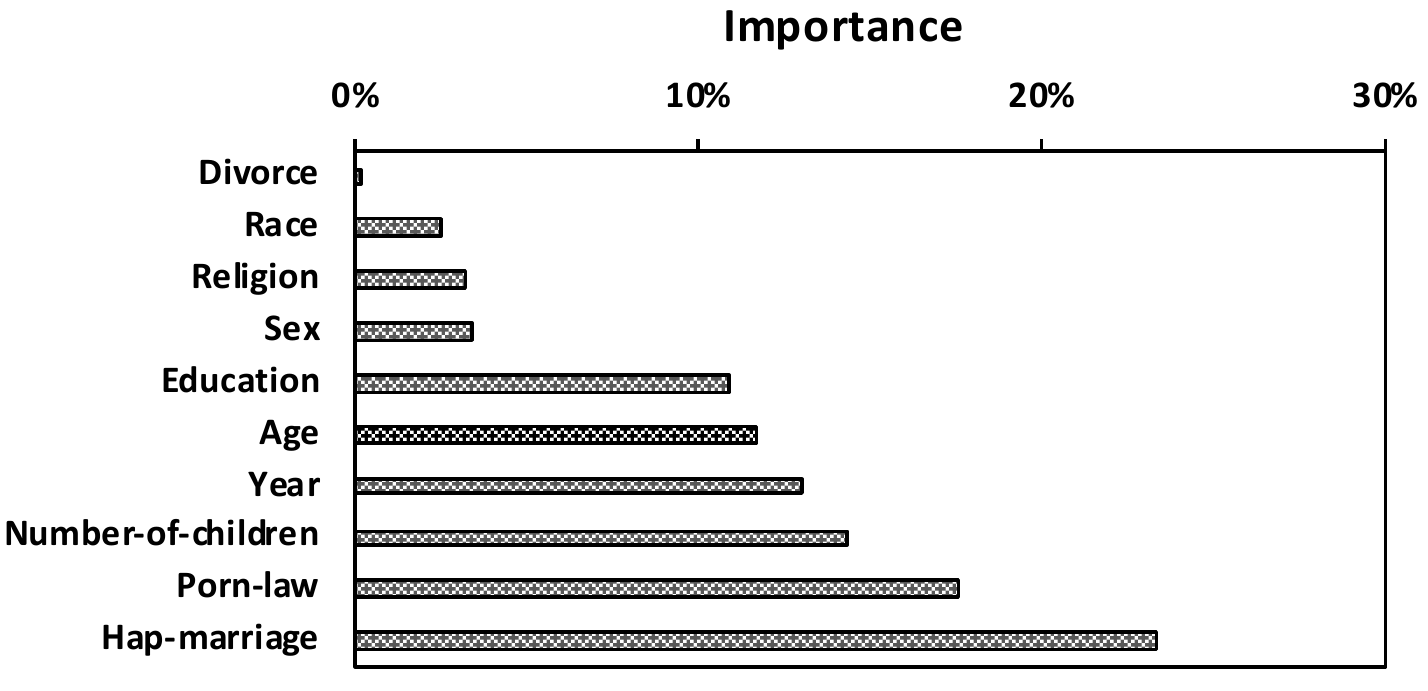}
}
\hfill
\subfigure[GSS dataset attributes' importance in the LOMIA attack model trained against the deepnet target model]{\includegraphics[width=0.45\textwidth, height=4cm]
{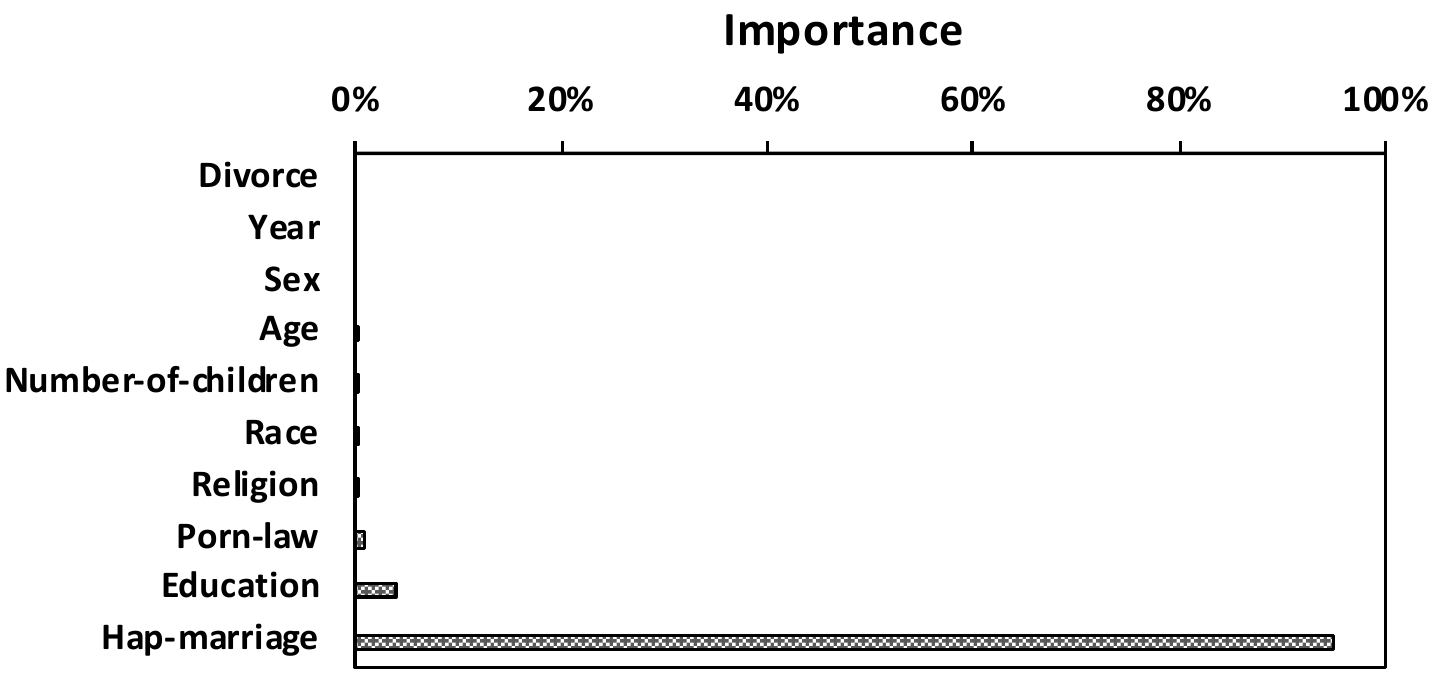}
}
\caption{{Importance of GSS dataset attributes in the LOMIA attack models trained against the decision tree and deepnet target models, respectively.}}
\label{fig:importance_gss}
\end{figure*}

\begin{figure*}[h]
\centering
\includegraphics[width=0.65\textwidth, height=4cm]
{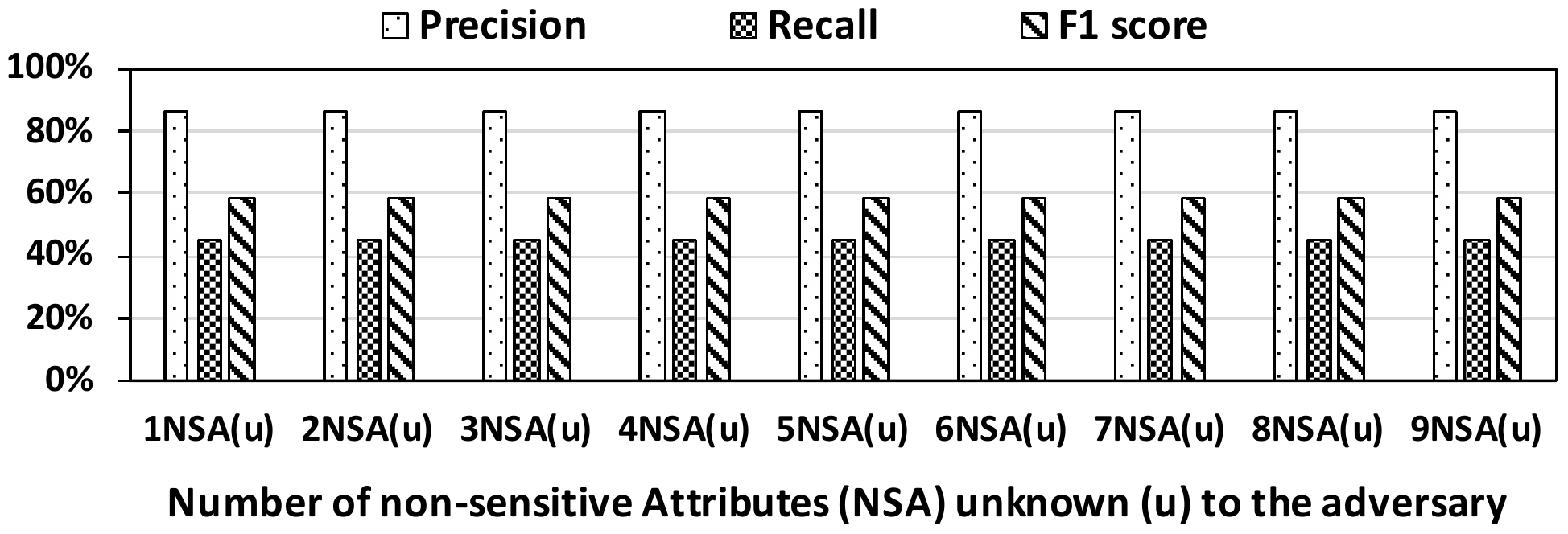}
\caption{{LOMIA performance against the deepnet model trained on Adult dataset when 1-9 non-sensitive attributes (NSA) increasingly become unknown (u) to the adversary in the following order: work-class, sex, race, fnlwgt, occupation, education, hours-per-week, capital-gain, and capital-loss. See Figure~\ref{fig:importance_adult} (b) for order.}
}
\label{fig:lomia_adult_dnn_partial}
\end{figure*}

\begin{figure*}[t]
\centering
\includegraphics[width=0.65\textwidth, height=4cm]
{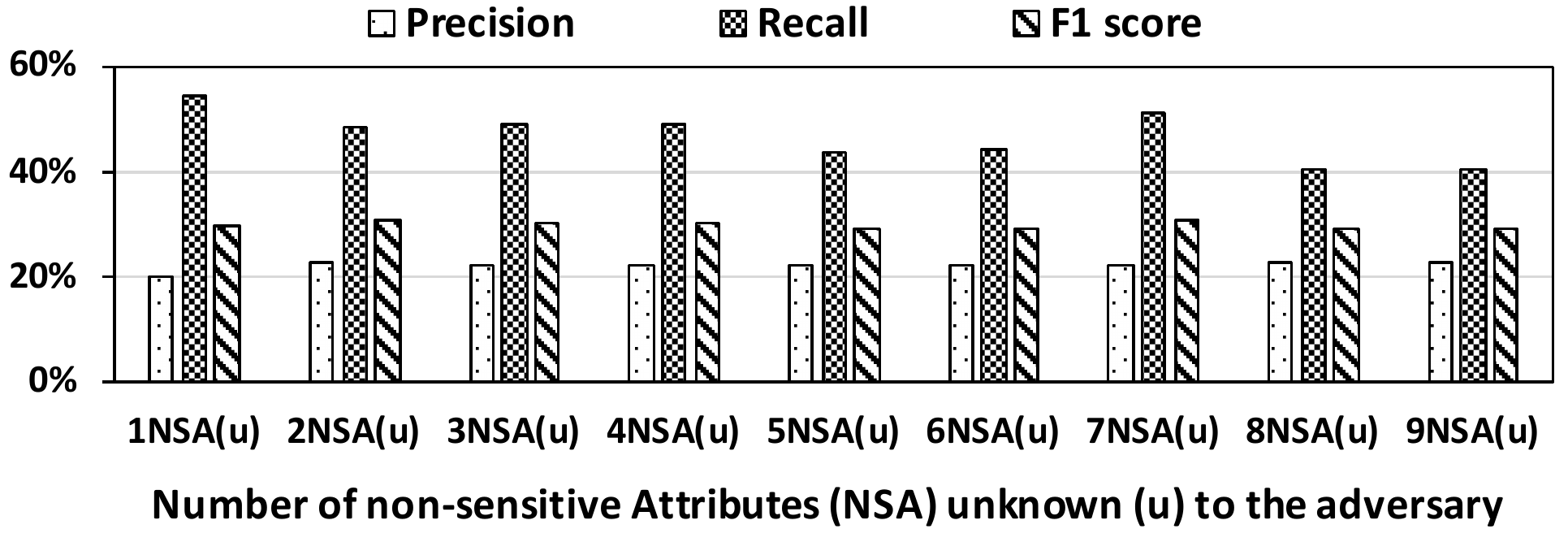}
\caption{{LOMIA performance against the decision tree model trained on GSS dataset when 1-9 non-sensitive attributes (NSA) are unknown (u) to the adversary in the following order: divorce, race, religion, sex, education, age, year, number-of-children, and porn-law. See Figure~\ref{fig:importance_gss} (a) for order.}
}
\label{fig:lomia_gss_dt_partial}
\end{figure*}

\begin{figure*}[t]
\centering
\includegraphics[width=0.65\textwidth, height=4cm]
{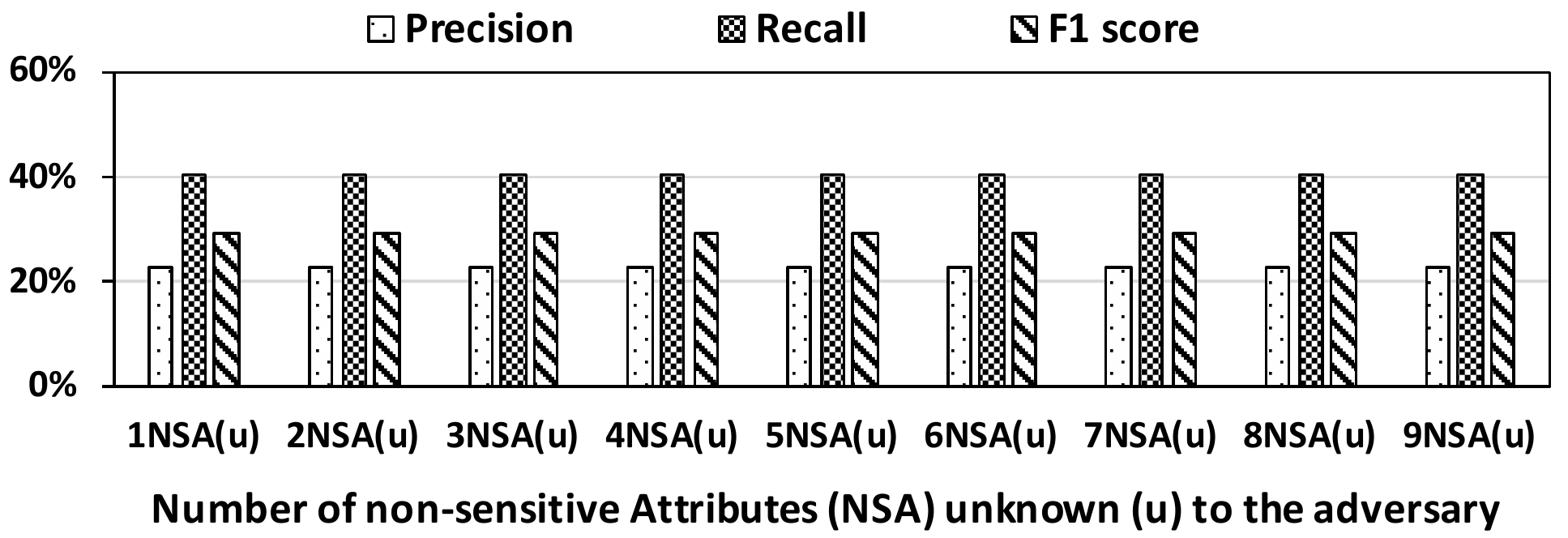}
\caption{{LOMIA performance against the deepnet model trained on GSS dataset when 1-9 non-sensitive attributes (NSA) are unknown (u) to the adversary in the following order: divorce, year, sex, age, number-of-children, race, religion, porn-law, and education. See Figure~\ref{fig:importance_gss} (b) for order.}
}
\label{fig:lomia_gss_dnn_partial}
\end{figure*}

\subsection{Number of Queries' Comparison Among Attacks }
\label{appn:queries}
We present the query numbers for different attacks on different datasets in Table~\ref{table:querycount}. For each attack experiment, all three attacks perform the same number of queries to the target model. GSS $DS_T$ dataset has 15235 instances and sensitive attribute \textit{x\_movie} has two possible values. Therefore, the  total number of queries for all attacks in this dataset is 15235x2=30470. The total number of queries for estimating the single sensitive attributes in Adult and FiveThirtyEight datasets are calculated similarly. For multiple sensitive attribute inference, i.e., estimating age group and alcohol in FiveThirtyEight dataset, we consider one sensitive attribute to be missing~\cite{bigmlLOMIA} and query the target model with all possible values of the other sensitive attribute. Therefore, the total number of queries while simultaneously estimating age group and alcohol sensitive attributes is 331x(4+2)=1986.

\begin{table*}[h]
\vspace{0.5cm}
  \centering
  \caption{Confusion matrix of FJRMIA on decision tree target model trained on FiveThirtyEight dataset (inferring multiple sensitive attributes: age and alcohol) }
    \vspace{0.2cm}
  \resizebox{0.75\textwidth}{!}{
\begin{tabular}{|l|*{6}{c|}}\hline
    \backslashbox{Actual}{Predicted}
    &18-29&30-44&45-60&>60& Total & Recall \\\hline
   18-29 & 0 & 64 & 0 & 6 & 70 & 0\% \\\hline 
    30-44  & 0 & 88 & 0 & 5 & 93 & 94.62\% \\\hline 
    45-60& 0 & 84 & 0 & 2 & 86 & 0\% \\\hline 
    >60& 0 & 77 & 0 & 5 & 82 & 6.1\% \\\hline 
   Total         & 0 & 313 & 0 & 18 & 331 & Avg. recall 25.18\% \\\hline 
    Precision     & 0\% & 28.12\% & 0\% & 27.78\% & Avg. precision 13.97\% & Accuracy 28.1\% \\\hline 

\end{tabular}
  }
\label{table:CM_DT_FTE_3_FJRMIA_app}

\vspace{0.3cm}

  \centering
  \caption{Confusion matrix of CSMIA on decision tree target model trained on FiveThirtyEight dataset (inferring multiple sensitive attributes: age and alcohol) }
    \vspace{0.2cm}
  \resizebox{0.75\textwidth}{!}{
\begin{tabular}{|l|*{6}{c|}}\hline
    \backslashbox{Actual}{Predicted}
    &18-29&30-44&45-60&>60& Total & Recall \\\hline
    18-29 & 35 & 12 & 7 & 16 & 70 & 0.5\% \\\hline 
    30-44  & 14 & 52 & 12 & 15 & 93 & 55.91\% \\\hline 
    45-60& 16 & 14 & 36 & 20 & 86 & 41.86\% \\\hline 
    >60& 16 & 24 & 17 & 25 & 82 & 30.49\% \\\hline 
   Total         & 81 & 102 & 72 & 76 & 331 & Avg. recall 44.57\% \\\hline 
    Precision     & 43.21\% & 50.98\% & 50\% & 32.89\% & Avg. precision 44.27\% & Accuracy 44.71\% \\\hline 

\end{tabular}
  }
\label{table:CM_DT_FTE_3_CSMIA_app}

\vspace{0.3cm}

  \centering
  \caption{Confusion matrix of LOMIA on decision tree target model trained on FiveThirtyEight dataset (inferring multiple sensitive attributes: age and alcohol) }
    \vspace{0.2cm}
  \resizebox{0.75\textwidth}{!}{
\begin{tabular}{|l|*{6}{c|}}\hline
    \backslashbox{Actual}{Predicted}
    &18-29&30-44&45-60&>60& Total & Recall \\\hline
    18-29 & 33 & 23 & 13 & 1 & 70 & 47.14\% \\\hline 
    30-44  & 24 & 48 & 15 & 6 & 93 & 51.61\% \\\hline 
    45-60& 19 & 29 & 33 & 5 & 86 & 38.37\% \\\hline 
    >60& 21 & 34 & 15 & 12 & 82 & 14.63\% \\\hline 
   Total         & 97 & 134 & 76 & 24 & 331 & Avg. recall 37.94\% \\\hline 
    Precision     & 34.02\% & 35.82\% & 43.42\% & 50\% & Avg. precision 40.82\% & Accuracy 38.07\% \\\hline  

\end{tabular}
  }
\label{table:CM_DT_FTE_3_LOMIA_app}

\vspace{0.3cm}

  \centering
  \caption{Confusion matrix of CSMIA (Case 1) on decision tree target model trained on FiveThirtyEight dataset (inferring multiple sensitive attributes: age and alcohol) }
    \vspace{0.2cm}
  \resizebox{0.75\textwidth}{!}{
\begin{tabular}{|l|*{6}{c|}}\hline
    \backslashbox{Actual}{Predicted}
    &18-29&30-44&45-60&>60& Total & Recall \\\hline
    18-29 & 16 & 0 & 0 & 0 & 16 & 100\% \\\hline 
    30-44  & 0 & 18 & 0 & 1 & 19 & 94.74\% \\\hline 
    45-60& 1 & 0 & 17 & 1 & 19 & 89.47\% \\\hline 
    >60& 1 & 1 & 0 & 6 & 8 & 75\% \\\hline 
    Total         & 18 & 19 & 17 & 8 & 62 & Avg. recall 89.8\% \\\hline 
    Precision     & 88.89\% & 94.74\% & 100\% & 75\% & Avg. precision 89.66\% & Accuracy 91.94\% \\\hline

\end{tabular}
  }
\label{table:CM_DT_FTE_3_CSMIA_case1_app}

\vspace{0.3cm}

  \centering
  \caption{Confusion matrix of LOMIA (Case 1) on decision tree target model trained on FiveThirtyEight dataset (inferring multiple sensitive attributes: age and alcohol) }
  \vspace{0.2cm}
  \resizebox{0.75\textwidth}{!}{
\begin{tabular}{|l|*{6}{c|}}\hline
    \backslashbox{Actual}{Predicted}
    &18-29&30-44&45-60&>60& Total & Recall \\\hline
    18-29 & 15 & 1 & 0 & 0 & 16 & 93.75\% \\\hline 
    30-44  & 0 & 18 & 0 & 1 & 19 & 94.74\% \\\hline 
    45-60& 1 & 0 & 16 & 2 & 19 & 84.21\% \\\hline 
    >60& 1 & 1 & 0 & 6 & 8 & 75\% \\\hline 
   Total         & 17 & 20 & 16 & 9 & 62 & Avg. recall 86.92\% \\\hline 
    Precision     & 88.24\% & 90\% & 100\% & 66.67\% & Avg. precision 86.23\% & Accuracy 88.71\% \\\hline 

\end{tabular}
  }
\label{table:CM_DT_FTE_3_LOMIA_case1_app}
\end{table*}

\begin{table*}[t]
  
  \centering
  \caption{Inferring the sensitive attribute alcohol, attack performances against the decision tree target model trained on FiveThirtyEight dataset (adversary also estimates the age group sensitive attribute).}
  \resizebox{.85\textwidth}{!}{
  \begin{tabular}{ | l | l | l | l | l | l | l | l | l | l | l |}
    \hline
    Attack Strategy &  TP & TN & FP & FN & Precision & Recall & Accuracy & F1 score & G-mean & MCC \\ \hline
    FJRMIA~\cite{FredriksonCCS2015} & $256$ & $5$ & $60$ & $10$ & $81.01\%$ & $96.24\%$ & $78.85\%$ & $87.97\%$ & $27.21\%$ & $7.51\%$\\ \hline 
    CSMIA & $151$ & $34$ & $31$ & $115$ & $82.97\%$ & $56.77\%$ & $55.89\%$ & $67.41\%$ & $54.49\%$ & $7.25\%$\\ \hline 
    LOMIA & $192$ & $19$ & $46$ & $74$ & $80.67\%$ & $72.18\%$ & $63.75\%$ & $76.19\%$ & $45.93\%$ & $1.25\%$ \\ \hline

  \end{tabular}
  }
\label{table:fte_exp3_results_dt}

\end{table*}

\begin{table*}[t]
  
  \centering
  \caption{Inferring the sensitive attribute alcohol, attack performances against the decision tree target model trained on FiveThirtyEight dataset}
  \resizebox{.85\textwidth}{!}{
  \begin{tabular}{ | l | l | l | l | l | l | l | l | l | l | l |}
    \hline
    Attack Strategy &  TP & TN & FP & FN & Precision & Recall & Accuracy & F1 score & G-mean & MCC \\ \hline
    FJRMIA~\cite{FredriksonCCS2015} & $266$ & $0$ & $65$ & $0$ & $80.36\%$ & $100.00\%$ & $80.36\%$ & $89.11\%$ & $0.00\%$ & $0.00\%$\\ \hline 
    CSMIA & $137$ & $40$ & $25$ & $129$ & $84.57\%$ & $51.50\%$ & $53.47\%$ & $64.02\%$ & $56.30\%$ & $10.36\%$\\ \hline 
    LOMIA & $198$ & $28$ & $37$ & $68$ & $84.26\%$ & $74.44\%$ & $68.28\%$ & $79.04\%$ & $56.63\%$ & $15.33\%$ \\ \hline

  \end{tabular}
  }
\label{table:fte_exp1_results_dt}

\end{table*}

\begin{table*}[t]

  \centering
  \caption{Query Numbers for Different Attacks}
  \resizebox{1\textwidth}{!}{
  \begin{tabular}{ | c | c | c | c |c |c |}
    \hline
    \multirow{2}{*}{Attack Strategy}  & GSS (x-movie),  &  Adult (marital-status),   & Fivethirtyeight (alcohol),  & Fivethirtyeight (age-group),  & Fivethirtyeight (age-group   \\
            & Section~\ref{subsubsec:expres_mia_gss}     & Section~\ref{subsubsec:expres_mia_adult}     & Section~\ref{subsec:attack_stability}   & Section~\ref{subsubsec:expres_mia_fte} (i)  & and alcohol), Section~\ref{subsubsec:expres_mia_fte} (ii) \\ \hline

   {FJRMIA}   & {30470} & {70444} & {662} & {1324} & {1986} \\
   \hline
   {CSMIA}  & {30470} & {70444} & {662} & {1324} & {1986}  \\
   \hline
   {LOMIA} & {30470} & {70444} & {662} & {1324} & {1986}  \\
    \hline
  \end{tabular}
  }
\label{table:querycount}
\end{table*}

 \begin{table*}[h]
  \centering
  \caption{Number of queries to target model for CSMIA partial knowledge attack on decision tree target model trained on Adult dataset (Figure~\ref{fig:missing_adult_dt_cs}).}
  \resizebox{1\columnwidth}{!}{
  \begin{tabular}{ | l | l | }
    \hline
Missing attribute(s) & Number of queries to target model \\ \hline
Occupation &	986216 \\ \hline
Capital-gain &	8453280 \\ \hline
Hours-per-week &	6762624 \\ \hline
Capital-loss &	6621736 \\ \hline
Education &	211332 \\ \hline
Work-class &	493108 \\ \hline
Race &	352220 \\ \hline
Sex	& 140888 \\ \hline
None &	70444 \\ \hline
Occupation, Capital-gain &	118345920 \\ \hline
Occupation, Hours-per-week &	94676736 \\ \hline
Occupation, Capital-loss & 92704304 \\ \hline

  \end{tabular}
  }
\label{table:CSMIA_queries}
\end{table*}